\documentclass[11pt,a4paper]{article}
\usepackage{jheppub}
\pdfoutput=1
\usepackage{amsfonts,amssymb}
\usepackage{amsmath}
\usepackage{epsfig, graphicx}

\newcommand{\be}{\begin{equation}}
\newcommand{\ee}{\end{equation}}
\newcommand{\bea}{\begin{eqnarray}}
\newcommand{\eea}{\end{eqnarray}}
\newcommand{\ba}{\begin{eqnarray}}
\newcommand{\ea}{\end{eqnarray}}

\newcommand{\eqn}[1]{(\ref{#1})}
\newcommand{\beq}{\begin{equation}}
\newcommand{\eeq}{\end{equation}}
\newcommand{\beqa}{\begin{eqnarray}}
\newcommand{\eeqa}{\end{eqnarray}}
\newcommand{\beqar}{\begin{eqnarray*}}
\newcommand{\eeqar}{\end{eqnarray*}}

\def\nc {N_\mt{c}}

\def\t6 {T_\mt{D6}}

\newcommand{\mt}[1]{\textrm{\tiny #1}}

\def\T11{{T}^{1,1}}
\def\bear{\begin{eqnarray}}
\def\eear{\end{eqnarray}}

\newcommand{\pa}{\partial}

\newcommand{\pasl}{\pa\kern-.55em /}

\newcommand{\ksl}{k\kern-.55em /}

\DeclareFixedFont{\xiiss}{OT1}{cmss}{m}{n}{12}
\DeclareFixedFont{\ixss}{OT1}{cmss}{m}{n}{9}
\DeclareFixedFont{\cmrnine}{OT1}{cmr}{m}{n}{9}

\newcommand{\CCs}{\hbox{\ixss C\kern-.4emI}}
\newcommand{\ZZs}{\hbox{\ixss Z\kern-.4emZ}}

\def\t{\tilde}

\newcommand{\ed}{{\cal E}}

\newcommand{\beal}[1]{\begin{eqnarray}\label{#1}}
\newcommand{\bel}[1]{\begin{equation}\label{#1}}

\newcommand{\bit}{\begin{itemize}}
\newcommand{\eit}{\end{itemize}}
\newcommand{\ben}{\begin{enumerate}}
\newcommand{\een}{\end{enumerate}}

\newcommand{\pl}{{\cal P_\mt{L}}}
\newcommand{\pt}{{\cal P_\mt{T}}}
\newcommand{\deltap}{\Delta {\cal P}}

\preprint{ICCUB-13-067; ITP-UU-13/09}
\title{Holographic isotropization linearized}

\author[a,1]{Michal P. Heller\note{On leave from: \emph{National Centre for Nuclear Research, Ho{\.z}a 69, 00-681 Warsaw, Poland}.}}
\author[b,c]{David Mateos}
\author[d]{Wilke van der Schee}
\author[a]{and Miquel Triana}

\affiliation[a]{Instituut voor Theoretische Fysica, Universiteit van Amsterdam,\\
Science Park 904, 1090 GL Amsterdam, The Netherlands}
\affiliation[b]{Instituci\'o Catalana de Recerca i Estudis Avan\c cats (ICREA), \\Passeig Llu\'\i s Companys 23, E-08010, Barcelona, Spain}
\affiliation[c]{Departament de F\'\i sica Fonamental,  Institut de Ci\`encies del Cosmos, \\Universitat de Barcelona, Mart\'{\i}  i Franqu\`es 1, E-08028 Barcelona, Spain}
\affiliation[d]{Institute for Theoretical Physics and Institute for Subatomic Physics,\\
Utrecht University, Leuvenlaan 4, 3584 CE Utrecht, The Netherlands}

\emailAdd{m.p.heller@uva.nl}
\emailAdd{dmateos@icrea.cat}
\emailAdd{W.vanderSchee@uu.nl}
\emailAdd{miquel.trianaiglesias@student.uva.nl}

\abstract{
The holographic isotropization of a highly anisotropic, homogeneous, strongly coupled, non-Abelian plasma was simplified in Ref.~\cite{Heller:2012km} by linearizing Einstein's equations around the final, equilibrium state. This approximation  reproduces the expectation value of the boundary stress tensor with a 20\% accuracy.  Here we elaborate on these results and extend them to observables that are directly sensitive to the bulk interior, focusing for simplicity on the entropy production on the event horizon. We also consider next-to-leading-order corrections and show that the leading terms alone provide a better description of the isotropization process for the states that are furthest from equilibrium.
}
\begin{document}  

\maketitle

%%%%%%%%%%%%%

\section{Introduction}

One of the most interesting recent developments in the gauge-gravity duality (holography) \cite{Maldacena:1997re,Witten:1998qj,Gubser:1998bc} is the study of thermalization processes in strongly coupled plasmas by means of their dual gravity description \cite{Chesler:2008hg,Chesler:2009cy,Chesler:2010bi,Heller:2011ju,Heller:2012je,Bantilan:2012vu,Heller:2012km,Schee:2013ab}.

Part of the motivation for these investigations comes from heavy ion collision (HIC) experiments at RHIC and LHC, in which the successful phenomenological description of data requires the applicability of viscous hydrodynamics with the quark-gluon plasma equation of state and a tiny shear viscosity $\eta/s = {\cal O}(1/4\pi)$ about $1$ fm/c or less after the collision \cite{Heinz:2004pj}. While this thermalization time is puzzlingly small from the  viewpoint  of bottom-up weak-coupling thermalization, it lies within the ball-park suggested by holographic calculations \cite{Chesler:2008hg,Chesler:2009cy,Chesler:2010bi,Heller:2011ju,Heller:2012je,Bantilan:2012vu,Heller:2012km,Schee:2013ab}. Although the dynamics in a real HIC surely involves a combination of weakly and strongly coupled  physics, the hope is that understanding the limit in which the physics is assumed to be strongly coupled at all length scales may help us bracket the real-world situation --- see \cite{CasalderreySolana:2011us} for a review of applications of holography to the QGP.

In the gauge-gravity duality the gravitational representation of the thermal state of a holographic quantum field theory in the planar limit and at strong ('t Hooft) coupling is a bulk black brane \cite{Witten:1998zw}. Thus thermalization on the gravity side is the process whereby a bulk geometry approaches the form of (a patch of) a static black brane and the other bulk fields relax to their thermal form in the black brane background. Generically, this stage is preceded by one which one may dub `hydrodynamization' whereby the system reaches a phase in which the system is still evolving but the dynamics is governed by hydrodynamic equations. On the gravity side this implies that the bulk geometry must have the form dictated by the so-called fluid-gravity duality \cite{Bhattacharyya:2008jc} 
(see e.g.~\cite{Hubeny:2011hd} for a review). As black brane metrics already depend on the bulk radial coordinate and we want to allow for at least time dependence, the studies of holographic thermalization processes require solving Einstein's equations in two or more variables, which in most cases requires the use of numerical relativity techniques.

Apart from toy-models based on the AdS-Vaidya geometry of infalling dust (see e.g. \cite{AbajoArrastia:2010yt,Balasubramanian:2010ce,Balasubramanian:2011ur,Aparicio:2011zy}, but also \cite{Bhattacharyya:2009uu}), the only existing approximation scheme to study holographic thermalization processes is the amplitude expansion, in which one linearizes Einstein's equations on top of the static black brane background. In this approximation the relaxation towards equilibrium is described by quasinormal modes with complex frequencies, whose imaginary parts lead to the damping of their amplitudes with time and hence to equilibration. These modes were thought so far to be appropriate for the description of only the late-time approach to equilibrium, when deviations from equilibrium are sufficiently small in amplitude \cite{Horowitz:1999jd}. 

An indication that this assumption might be too restrictive comes from black hole mergers in asymptotically flat four-dimensional spacetime. There, in the so-called close-limit approximation, the Einstein's equations linearized on top of the final black hole predict rather accurately the pattern of gravitational radiation at infinity provided the initial data have a single horizon surrounding the merging black holes \cite{Price:1994pm,Anninos:1995vf}. This initial horizon, however, needs not to be a small perturbation of the final black hole for the close-limit approximation to work.

These features, together with the observation that the AdS analogue of gravitational radiation at infinity is the expectation value of the boundary stress tensor, motivated recent work \cite{Heller:2012km} in which a linear approximation was applied to a simple example of far-from-equilibrium gravitational dynamics in AdS spacetime. Specifically, this reference considered the isotropization of a large number of initially homogeneous  but anisotropic states in a Conformal Field Theory with a holographic description, or holographic Conformal Field Theory (hCFT) for short, with a large number of colors $N_{c}$ and strong 't Hooft coupling \cite{'tHooft:1973jz} $\lambda = g_\mt{YM}^2 \nc$. The outcome of this analysis was that the linearized Einstein's equations indeed reproduce well the dynamics of the one-point function of the stress tensor. For small perturbations this (expectedly) provided excellent approximations to the time evolution of the stress tensor. Surprisingly, however, even for initial states which were not a small perturbation of the final state, the linearized evolution still predicted the stress-tensor with an accuracy of about $20\%$. The purpose of the present paper is to explore the applicability of the linearized Einstein's equations further, in particular for observables that are directly sensitive to the interior bulk dynamics such as the entropy production. 

We start this article by reviewing the results of \cite{Heller:2012km}. The key technical novelty of \cite{Heller:2012km}, compared to earlier works, was the ability to generate and follow the evolution of a large number (around 1000) of different non-equilibrium initial states. In the current article we provide a detailed explanation of how these initial states were obtained and evolved. We also comment on the relation between features of the initial states and the time it takes them to thermalize. 

Subsequently, we review the linearized gravity approach to holographic isotropization and comment on the relation between the form of initial data and the accuracy of the linear approximation. We also study in detail the decomposition of the solutions into quasinormal modes. According to the common lore, quasinormal modes can only describe the late-time approach to equilibrium, for which only the lowest-lying modes are relevant. As already anticipated in \cite{Heller:2012km}, our finding is that, provided a sufficient number of modes is considered, the description in terms of quasinormal modes can actually be extended all the way back to early times for which the system is very far from equilibrium. The correct description of the system \emph{at all times} then results from the heavy interference of the quasinormal modes with one another. 

Finally, we discuss the corrections to the linearized Einstein's equations.
For generic observables that are sensitive to the geometry deep in the bulk these corrections are crucial because they constitute the leading-order result. This is the case, for example, for the entropy production, on which we focus our attention.

Throughout our paper we will assume that the boundary stress tensor is diagonal. A systematic discussion of the amplitude expansion of Einstein's equations in the more general homogeneous but non-diagonal situation can be found in \cite{Mukhopadhyay:2012hv}.

\section{Holographic setup \label{sec.holosetup}}

\subsection{Holographic isotropization and the ansatz for the dual metric}

The holographic isotropization that we will study is the dynamics of a hCFT, in which a strongly coupled $(3+1)$-dimensional plasma is homogeneous, but has a time-dependent pressure anisotropy leading to rich non-equilibrium physics \cite{Chesler:2008hg,Heller:2012km}. In this setup all the local operators apart from the stress tensor have vanishing expectation values, which constitutes the universal sector of any hCFT \cite{Bhattacharyya:2008mz}. 

This example of a holographic thermalization process is, most likely, the simplest one, as due to homogeneity it does not involve hydrodynamic modes and thermalization is achieved directly via relaxation to a thermal state without being preceded by a long hydrodynamic phase. The latter observation is the primary reason for considering it here, and previously in \cite{Heller:2012km}, in the context of the close-limit approximation for anti-de Sitter spacetimes.

As in \cite{Heller:2012km}, we will be interested in an unforced relaxation to the thermal state and we consider the situation in which there are no time-dependent sources pumping energy into our system, i.e.~the boundary metric is flat. This is the crucial technical difference between our work and that of  reference \cite{Chesler:2008hg}, which considered the holographic isotropization process in which non-equilibrium states were obtained by placing the boundary theory in a time-dependent metric for a finite period of time.

The form of the bulk metric is determined to a large extent by the ansatz for the boundary stress tensor. After imposing for simplicity rotational symmetry in two of the spacelike directions and taking into account the flatness of the boundary metric, the most general conserved and traceless stress tensor can be written as
\be
\langle T_{\mu \nu} \rangle = \frac{\nc^2}{2\pi^2}
\mathrm{diag} \Big[ \ed, \, \pl(t), \, \pt(t), \, \pt(t) \Big] \,,
\ee
where $\nc$ is the rank of the $SU(\nc)$ gauge group and
$\ed$ is proportional to the energy density. The longitudinal $\pl(t)$ and transverse pressures $\pt(t)$ are then  expressed in terms of a time-dependent anisotropy $\deltap(t)$ through
\bea
\pl(t) &=& \tfrac{1}{3} \ed - \tfrac{2}{3} \deltap (t) \, ,\\ [2mm]
\pt (t) &=& \tfrac{1}{3} \ed + \tfrac{1}{3} \deltap (t) \, .  
\eea
Note that the energy density in our setup is constant in time by virtue of the homogeneity assumption plus the conservation of the stress tensor. In this sense the energy density is part of the initial conditions. As the only possible static state with finite energy density  is the isotropic and homogeneous plasma \cite{Janik:2008tc}, the final state is known already from the start, without the need to solve any dynamical equation. This seems to be a rather non-generic feature of our setup, which we discuss at length in the conclusions section.

The symmetries of the stress tensor  -- through the general near-boundary form of the metric in the Fefferman-Graham coordinates \cite{deHaro:2000xn} -- suggest the ansatz for the dual metric $g_{a b}$ with $g_{tt}$, $g_\mt{LL}$ and $g_\mt{TT}$ components being dynamical, i.e. obtained by solving the equations of motion.\footnote{It does not have to be necessarily the case and actually in the ADM formulation of general relativity it is the form of the lapse -- the timelike warp factor ($g_{tt}$) -- and the shift ($g_{t a}$) that are fixed/imposed. For details in the context of AdS spacetimes see \cite{Heller:2012je}.} The freedom of defining what is meant by the bulk time $t$ and the radial coordinate in AdS $r$ has to be then (partly) fixed by imposing the form of the $g_{rr}$ and $g_{tr}$ components.

In our analysis it will be very convenient to follow \cite{Chesler:2008hg} and express the dual metric $g_{a b}$ in the generalized ingoing Eddington-Finkelstein coordinates
\begin{equation}
\label{mansatz}
ds^2 = 2 dt dr - A dt^{2} + \Sigma^{2} e^{-2 B} dx_\mt{L}^{2} + \Sigma^{2}e^{B} d\mathrm{\mathbf{x}}_\mt{T}^{2} \,,
\end{equation}
where $A$, $\Sigma$ and $B$ are functions of time $t$ and the radial coordinate $r$. With this ansatz $g_{rr} = 0$ and $g_{t r} = 1$. In the ingoing Eddington-Finkelstein coordinates, constant time slices are null hypersurfaces as radial ingoing null geodesics, by construction, penetrate the bulk instantaneously.

The metric \eqref{mansatz} has to solve the Einstein's equations with the negative cosmological constant
\be
\label{EinsteinEQNS}
R_{a b} - \frac{1}{2} R \, g_{a b} - \frac{6}{L^2} \, g_{a b} = 0,
\ee
where in the following we set the radius of the AdS solution $L$ \mbox{to $1$}. The link between the form of the field theory stress tensor and the dual metric ansatz becomes clear after solving Einstein's equations in the near-boundary (large $r$) expansion 
\begin{subequations}
\label{nearbdryexpansions}
\begin{eqnarray}
\label{nearbdryexpansions.a}
A&=& r^2 + \frac{a_{4} }{r^{2}} - \frac{2b_{4}(t)^{2} }{7 r^{6}}  + \cdots \,, \\ [2mm]
\label{nearbdryexpansions.b}
B&=&\frac{b_{4}(t)}{r^4} + \frac{\partial_{t} b_{4}(t)}{r^5}  +  \cdots  \,, \\ [2mm]
\label{nearbdryexpansions.c}
\Sigma &=& r - \frac{b_{4}(t)^{2}}{7 r^{7}}  + \cdots \,, \,\,\,\,\,\,\,
\end{eqnarray}
\end{subequations}
and identifying normalizable modes with the components of the stress tensor using  holographic renormalization \cite{deHaro:2000xn}. For the specific case of $SU(\nc)$ ${\cal N}=4$ super Yang-Mills theory this relation is  
\be
\ed = - 3 a_{4}/4 \quad \mathrm{and} \quad 
\deltap (t) = 3 b_{4}(t) \,.
\label{related}
\ee
The metric ansatz \eqref{mansatz} leaves the residual diffeomorphisms freedom $r \rightarrow r + f(t)$. This freedom is fixed here by imposing the absence of the term proportional to $r$ in the near-boundary expansion for the $g_{tt}$ metric component \eqref{nearbdryexpansions.a}.

In \eqref{nearbdryexpansions} we suppressed the near-boundary expansion at relatively low order, but it is important to stress that the expansion has infinitely many terms carrying arbitrarily high derivatives of the pressure anisotropy. This inevitably leads to a general conclusion that a state given by the form of the geometry on a constant time slice is (partly) specified by infinitely many derivatives of the dual stress tensor, in our case the pressure anisotropy.

\subsection{Thermalization criterium}

Although $\ed$ is constant in time, the physical temperature can only be assigned to the system once the equilibrium is reached. In this regime  $\ed=3\pi^4 T^4/4$ and the metric takes the form
\be
\label{eq.AdSSchwarzschildLineElem}
ds^{2} = 2 dt dr - r^2 \left(1 - \frac{(\pi T)^{4}}{r^{4}}\right) dt^{2} + r^{2} d\vec{x}^{2} \,,
\ee
or in terms of $A$, $\Sigma$ and $B$
\be
\label{eq.AdSSchwarzschildComponents}
A = r^{2} \left(1 - \frac{(\pi T)^{4}}{r^{4}}\right), \quad \Sigma = r \quad \mathrm{and} \quad B = 0,
\ee
and describes the AdS-Schwarzschild solution between the boundary and the future event horizon covering also the black brane interior.

Although equilibration of a holographic system can be sampled with different field theory probes, including expectation values of local operators, two point functions, entanglement entropy and Wilson loops, in our studies we will primarily focus on tracing the evolution of the one-point function of the stress tensor. There are two reasons for this. In the first place this is the quantity of interest if one wants to make a phenomenological contact with the  fast applicability of hydrodynamics at RHIC and LHC. Secondly, after the stress tensor eventually settles down to its thermal value, the geometry becomes a patch of the AdS-Schwarzschild black brane and from this moment on there is no need to evolve the Einstein's equations further.

We will hence define thermalization time as the isotropization time $t_\mt{iso}$, i.e.~the time after which the stress tensor anisotropy $\deltap (t)$ remains small compared to the energy density and eventually decays to zero. In our calculations, as in \cite{Heller:2012km}, we adopt the following criterium for $t_\mt{iso}$:
\be
\label{eq.tiso}
\deltap(t > t_\mt{iso}) \leq 0.1 \, \ed,
\ee
but it is important to keep in mind that thermalization is never a clean-cut event and the threshold on the RHS of \eqref{eq.tiso} can be always slightly raised or lowered without altering much the results.

\subsection{The event horizon and its entropy}

The event horizon is defined as the causal boundary of the black hole. It is a teleological object which can be located only after the black hole settles down to the state of permanent equilibrium. 

For the purposes of the current paper, we will be interested in the event horizon's area as an example of one easy-to-compute bulk observable that is directly sensitive to the form of the geometry in the deep IR. Although no precise definition of the entropy density exists in a truly far-from-equilibrium situation, the change in the area density of the event horizon  provides a crude measure of the total entropy produced in the thermalization process. For this reason we will loosely refer to the area density of the event horizon as `entropy density', 
but we emphasize from the start that this terminology is rigorously applicable only near equilibrium. We also clarify that we chose to focus on the event horizon, as opposed to e.g.~the apparent horizon, because in many cases there was no apparent horizon on our initial-time slice.

Due to the symmetries of our problem, the event horizon will be a hypersurface defined by
\be
r - r_\mt{EH}(t) = 0,
\ee
with the normal vector being null
\be
\label{eq.horloc}
r_\mt{EH}'(t) - \frac{1}{2} A\left(t, r_\mt{EH}(t)\right) = 0.
\ee
The latter is the geodesic equation for the outgoing light ray and needs to be supplemented with the following condition to be imposed in the asymptotic future
\be
r_\mt{EH}(t) \rightarrow \pi T \quad \mathrm{as} \quad t \rightarrow \infty \,.
\ee
In practical terms this condition implies that when the metric eventually approaches the form of the AdS-Schwarzschild black brane \eqref{eq.AdSSchwarzschildLineElem}, $r_\mt{EH}$ approaches the position of the event horizon of the static solution.

The area of the event horizon gives rise to the following expression for the entropy
\be
\label{eq.entropy}
s_\mt{EH}(t) = \frac{1}{2 \pi}N_{c}^{2} \, \Sigma\left(t,r_\mt{EH}(t)\right)^{3},
\ee
which is guaranteed to be a non-decreasing function of $t$. In \eqref{eq.entropy} we implicitly assumed that the event horizon's area is mapped onto the boundary along ingoing null radial geodesics, i.e.~along lines of constant $t$ for the metric ansatz \eqref{mansatz}. 
%This is a reasonable choice  in our setup due to its high degree of symmetry.

\section{Far-from-equilibrium initial states and their nonlinear evolution \label{sec.nonlinearEEQ}}
\subsection{Solving Einstein's equations as an initial-value problem}

As originally noted in \cite{Chesler:2008hg}, the Einstein's equations for the metric ansatz \eqref{mansatz} can be neatly written as
\begin{subequations}
\begin{eqnarray}
\label{Seq}
0 &=& \Sigma \, (\dot \Sigma)' + 2 \Sigma' \, \dot \Sigma - 2 \Sigma^2\,,
\\ \label{Beq}
0 &=& \Sigma \, (\dot B)' + {\textstyle \frac{3}{2}}
    \big ( \Sigma' \dot B + B' \, \dot \Sigma \big )\,,
\\  \label{Aeq}
0 &=& A'' + 3 B' \dot B - 12 \Sigma' \, \dot \Sigma/\Sigma^2 + 4\,,
\\  \label{Cr}
0 &= & \ddot \Sigma
    + {\textstyle \frac{1}{2}} \big( \dot B^2 \, \Sigma - A' \, \dot \Sigma \big)\,,
\\ \label{Ct}
0 &=& \Sigma'' + {\textstyle \frac{1}{2}} B'^2 \, \Sigma\,,
\end{eqnarray}
\label{Eeqns}
\end{subequations}
where 
\begin{equation}
\label{eq.defdirderivs}
h' \equiv \partial_r h \quad \mathrm{and} \quad \dot h \equiv \partial_t h + {\textstyle \frac{1}{2}} A \, \partial_r h
\end{equation}
denote respectively derivatives along the ingoing and outgoing radial null geodesics. In the following we will be interested in solving the initial-value problem, i.e.~given the geometry on the initial-time slice we want to obtain the evolution of the dual stress tensor by computing the bulk spacetime outside the event horizon.

Not all equations among \eqref{Eeqns} are evolution equations, i.e.~specify the form of the metric on a neighboring time slice. Equations \eqref{Cr} and \eqref{Ct} are constraints in the sense that the remaining components of the Einstein's equations can be shown to guarantee that they are obeyed provided \eqref{Cr} holds at the boundary and \eqref{Ct} holds on the initial-time slice \cite{Chesler:2008hg}. 

The null character of the coordinate system \eqref{mansatz} leads to a nested algorithm for solving the initial-value problem in which one uses as evolution equations \eqref{Seq}-\eqref{Aeq} and at each time step one only needs to solve linear ordinary differential equations in $r$. The precise scheme that we will follow is a slight modification of the one originally introduced in \cite{Chesler:2008hg}, and consists of the following steps:
\begin{enumerate}
\item we start with $B$ as a function of $r$ and the energy density $\ed$ (constant in our setup);
\item the constraint equation \eqref{Ct} allows us to solve for $\Sigma$ as a function of $r$;
\item we then solve \eqref{Seq} for $\dot{\Sigma}$, with $\ed$ being the integration constant;
\item having $B$, $\Sigma$ and $\dot{\Sigma}$, we solve \eqref{Beq} for $\dot{B}$;
\item with $B$, $\Sigma$, $\dot{B}$ and $\dot{\Sigma}$ at hand we can integrate \eqref{Aeq} for $A$;
\item knowing $\dot{B}$ and $A$ and using \eqref{eq.defdirderivs} we get $\partial_t B$;
\item we proceed to the next time step using a finite difference scheme (for details see section \ref{subsec.numerics}).
\end{enumerate}
In our setup the constraint \eqref{Cr} is implemented in the near-boundary expansion, as it encodes the conservation of the stress tensor in the dual gauge theory. In order to monitor the accuracy of the numerical code we check the value of this constraint in the bulk when evaluated on the numerical solution (see also section \ref{subsec.numerics}).

The algorithm outlined above needs to be supplemented with the initial conditions, the choice of which we discuss in the next subsection. 

\subsection{Specifying initial states \label{sec.specinistates}}

Gravity encodes dual initial states in the form of the geometry on a bulk initial-time slice. The conditions on the initial data arise from three sources: the constraint \eqref{Ct}, the near-boundary expansion \eqref{nearbdryexpansions} and bulk regularity. By the latter we mean that all possible singularities in the initial data must be hidden inside the event horizon. 

One way to obtain a non-equilibrium state while automatically satisfying the conditions above is to start with vacuum AdS and perturb it by turning on a non-normalizable mode of the bulk metric or some other bulk field for a finite period of time \cite{Chesler:2008hg}. The alternative approach, that we adopt here and which was used also in \cite{Beuf:2009cx,Heller:2011ju,Heller:2012je}, is to start with non-equilibrium states defined as solutions of the constraints on the initial-time slice without invoking the way in which a particular state was created.

Equation \eqref{Ct} imposes a constraint between the forms of $B$ and $\Sigma$ on the initial-time slice. Since $B$ appears quadratically, we choose to specify the initial state through $B$ and then use \eqref{Ct} to solve for $\Sigma$. Note that this equation, together with the asymptotic behavior linear in $r$ \eqref{nearbdryexpansions.c}, implies that $\Sigma$ must necessarily be a convex function and hence that it must vanish for some $r \geq 0$,  with the inequality being saturated only for vacuum AdS and the AdS-Schwarzschild black brane. As our coordinate frame is spanned by the ingoing radial null geodesics and $\Sigma$ measures the transverse area of the congruence, $\Sigma = 0$ implies reaching a caustic and hence the breakdown of our coordinate frame. 

For the \emph{successful} evolution of the initial data specified by some $B$ we thus need to make sure that the locus where $\Sigma$ vanishes is hidden behind the event horizon on the initial-time slice. As the event horizon is a teleological object, this cannot be verified a priori -- we need to try to run a simulation and when it is successful we know that the initial state we started with was legitimate. 

The contrary is not necessarily the case, as a caustic, a priori, is just a breakdown of a coordinate system. However, we verified numerically that in the neighborhood of a point where $\Sigma$ vanishes we obtain very large curvatures. This suggests that this point  \emph{must} be hidden inside the event horizon.\footnote{A possible caveat is that our numerics is not guaranteed to give trustable results inside the event horizon, as the physics might be influenced there by the artifacts of the inner cut-off of the grid.}

We thus see there is an interesting interplay between the choice of $B$ and the choice of the (initial) energy density $\ed$. Both quantities, a priori, seem to be very much independent when it comes to specifying the initial state. If, however, the point where $\Sigma$ vanishes corresponds to a genuine curvature singularity, which is the case for the AdS-Schwarzschild black brane and which our numerical studies also indicate, then there must be a minimal energy density $\ed$ for which this singularity is still covered by the event horizon on the initial-time slice. An indication that this might be the case comes from noting that for the AdS-Schwarzschild black brane the radial position of the event horizon is proportional to (the fourth root of) the energy density $\ed$. Similarly,  in our setup we can always put the final event horizon at a smaller value of $r$ than the point at which $\Sigma$ vanished on the initial-time slice by decreasing the energy density. This discussion suggests that the states for which the initial position of the event horizon is very close to the point where $\Sigma=0$ should be viewed as  \emph{maximally far from equilibrium}. 

In our setup, we have a freedom of preparing arbitrary initial conditions, i.e.~ we can specify $B$ as a function of $r$ on the initial-time slice and $\ed > 0$, as long as $B$ obeys the near-boundary expansion of the form \eqref{nearbdryexpansions.b} and there are no naked singularities. We use this freedom to prepare and follow the evolution of states in which $B$ has support very close to the boundary, very close to the horizon or spreads over a large range of the radial direction. In order to generate a large number of non-equilibrium initial states  we followed the following procedure:
\begin{enumerate}
\item without  loss of generality we choose units such that $a_{4} = -1$, or equivalently $\ed = \frac{3}{4}$;
\item we generate the initial $B$ as a ratio of two 10th order polynomials in $r$ with random coefficients in the range $(0,1)$;
\item we subtract from it a cubic expression so that the near-boundary expansion for $B$ of the form \eqref{nearbdryexpansions.b} is obeyed;
\item the whole expression is then normalized so that the maximal value of the $B$ between the boundary and the position of the final event horizon ($r = 1$) is $\frac{1}{2}$; 
\item we then run a binary search algorithm to find the factor that $B$ needs to be multiplied by such that the code is just stable, while storing successful runs. Typically, we repeat this step about 6 times per seed function generated in step 2.
\end{enumerate}

In this way we can generate states which are as far from equilibrium as our numerical code allows. In the end this means there is some sensitivity to the number of grid points, since increasing the number of grid points would improve the stability.

Finally, it is interesting to note that a constraint of exactly the  form \eqref{Ct}  also holds for metric ansatzes corresponding to a dual plasma expanding in one dimension \cite{Chesler:2009cy,Chesler:2010bi}. This implies that our discussion  about the specification of the initial states, including the maximally far from equilibrium ones, also applies in these other setups. However, if we relax the assumption of a homogeneity in the transverse plane, then $\Sigma$ is no longer forced to be convex and there might be bulk states which do not lead to caustics/apparent singularities in the way described above.

\subsection{Numerical implementation \label{subsec.numerics}}

In the numerical implementation instead of the variable $r$ we used its inverse
\be
z = \frac{1}{r},
\ee
so that the AdS boundary is at $z = 0$. With the choice of units we made in our code, i.e.~$a_{4} = -1$, the horizon of the final black brane is then located at $z = 1$. Note however that, for definiteness, all dimensionful quantities that we will provide will be specified in terms of the energy density or, equivalently, the temperature of the final black brane, which is the only scale at the moment of thermalization.

For discretization we use the pseudo-spectral collocation method for the spatial grid in the $z$ direction (see \cite{Wu:2011ab,lrr-2009-1} for accessible reviews of the spectral methods in the context of numerical GR). This allows us to maintain a very moderate grid, typically with 26 points, without a significant loss of accuracy. Each of the evolution equations \eqref{Seq}-\eqref{Aeq} is written in terms of redefined variables in which the three terms of lowest order in the near-boundary expansion are subtracted, and solved as a set of linear equations.

The latter is a significant simplification, since  it allows the  use of standard procedures for solving linear equations. We implemented our numerical code in \emph{Mathematica} and used the fourth order Adams-Bashforth method for evolution in time. The first five time steps were made with the use of low order Runge-Kutta methods. Typically, we used time steps of size $0.0025 \sim (\#\,\, \mathrm{grid}\,\, \mathrm{points})^{-2}$.

\begin{figure}
\begin{centering}
\includegraphics[width=7.5cm]{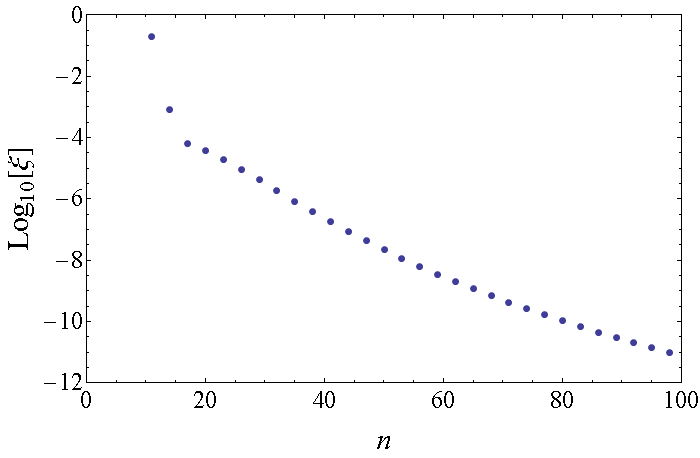}\includegraphics[width=7.5cm]{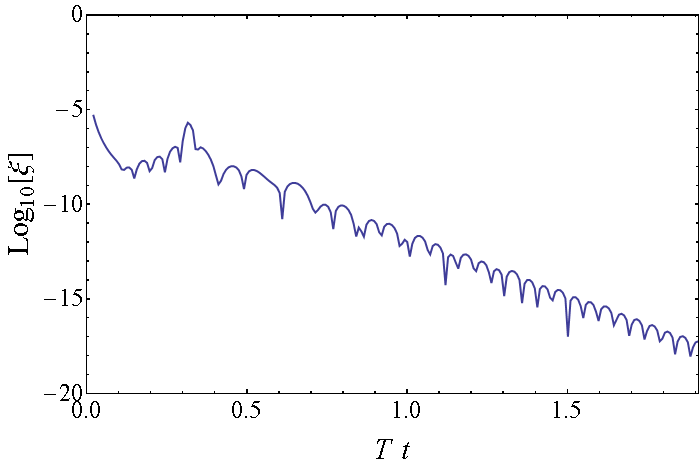}
\par\end{centering}
\caption{The left plot shows the value of the normalized constraint $\xi(t = 0)$ as a function of the number of grid points $n$ for the evolution of the initial profile $B(z,\, t_\mt{ini})= \frac{8}{3} {\cal E} z^{4}$. It is clear from the plot that our numerics converges exponentially with the number of grid points. The right plot shows the evolution of $\xi(t)$ as a function
of time for $n=26$ and one can see there that the constraint actually decreases with time. To achieve $\xi(t)<10^{-9}$ one typically
needs higher precision than the standard double precision computations offer.
\label{fig:constraint}}
\end{figure}

As a way of monitoring the accuracy of our code, we used the normalized \mbox{constraint \eqref{Cr}}
\be
\label{normalizedconstraint}
\xi(t)  = \mathrm{max}_{\,r} \left(\frac{\ddot{\Sigma} + \frac{1}{2} \dot{B}^{2} \Sigma - \frac{1}{2} A' \dot{\Sigma}}{|\ddot{\Sigma}| + \frac{1}{2} \dot{B}^2 \Sigma + \frac{1}{2} | A' \dot{\Sigma}|}\right){\Bigg |}_{\mathrm{fixed}\,\,t}
\ee
The convergence of our code is then illustrated by Fig.~\ref{fig:constraint}, which shows typical plots of the maximum value of the normalized constraint $\xi(t)$. 

The last feature that needs to be discussed is the choice of the position of the inner boundary of the computational grid. Note that the simulation is well-defined only if the grid covers the entire portion of the spacetime outside the event horizon. Initially this is hard to predict, since the position of the event horizon depends on the future evolution. Therefore one typically focuses attention on the presence of the apparent horizon because, if it can be found, it is guaranteed to lie inside the black hole. However, quite frequently in our case there was no apparent horizon on the initial-time slice and therefore we used the following procedure. We first tried to run simulations with the radial cut-off put at $z = 1.01$, which is right below the late-time position of the event horizon. This often worked,  and when it did not we reran the simulation with $z = 1.07$ as a cut-off. The latter point turned out to almost always lie past the event horizon. In this way we could successfully evolve a large number of initial states. 

\subsection{Why the near-boundary expansion is not enough}

The near-boundary expansion of the metric contains information about all time derivatives of the pressure anisotropy at the initial time. This allows the construction of a Taylor-series expansion for the pressure anisotropy, which at least for some time should be very close to the exact expression. Previous studies in the context of boost-invariant flow suggest that the convergence radius of a series of this type is too small to see thermalization \cite{Heller:2012je}. We can convince ourselves that this is the case also here by considering the initial profile of the form
\be
\label{constDP}
B_\mt{ini}(z) = \frac{8}{3} {\cal E} z^{4}. 
\ee
According to the near-boundary expansion \eqref{nearbdryexpansions.b} this profile at the initial time has a non-zero pressure anisotropy, but \emph{all} its time derivatives vanish. The Taylor series expression predicts then a constant pressure anisotropy, which would lead to a singular spacetime \cite{Janik:2008tc}.

\begin{figure}
\begin{centering}
\includegraphics[width=7cm]{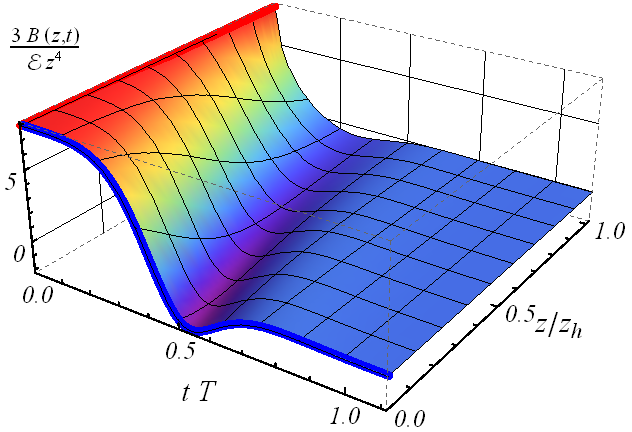}
\includegraphics[width=7cm]{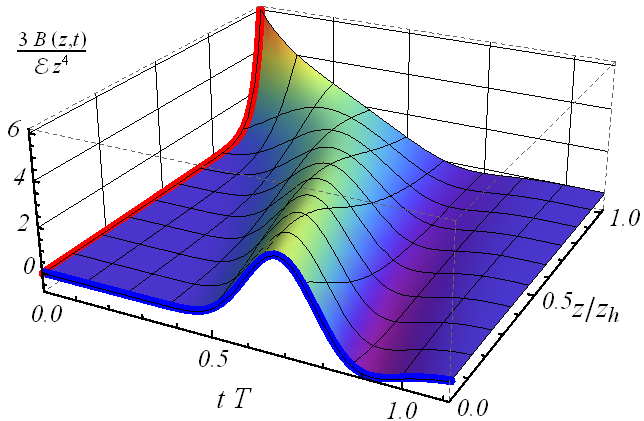}

\par\end{centering}
\caption{The left plot shows $B(z,\,t)$ for the initial profile \eqref{constDP}, which is shown as a thick red line at $t=0$. By construction, all time derivatives of the pressure anisotropy vanish at $t=0$. The thick blue curve at $z = 0$ shows the value of the gauge theory quantity $\deltap(t) / {\cal E}$. One sees clearly that the Taylor expansion around the initial time (predicting constant pressure anisotropy) does not converge when thermalization is achieved. \newline
The right plot, in which we start with $B(t = 0, z)=(\frac{4}{3}{\cal E})^6 z^{24}$, shows similar behavior. The initial disturbance, which is localized in the IR part of the geometry, propagates to the boundary in a time limited by causality. This creates the pressure anisotropy, which quickly relaxes back to zero.}
\label{Bexampderivs}
\end{figure}

Fig.~\ref{Bexampderivs} shows the numerical evolution of the profile \eqref{constDP} and proves that the radius of convergence of the Taylor expansion is indeed too small to see thermalization defined via \eqref{eq.tiso}. This result implies that the precise form of the metric $g_{a b}$ and the pressure anisotropy $\deltap (t) $ at transient times need to be obtained via an explicit solution of the initial-value problem on the gravity side.

\subsection{Evolution of a sample profile and expectations for thermalization times}

To get  intuition about how the dynamics proceeds on the gravity side and to get acquainted with the features following from the choice of a foliation by null constant-time slices, it will be instructive to discuss in detail the dynamics of the following initial state
\be
\label{sampleprofile}
B(t = 0, z)=\frac{2}{15} {\cal E} z^{4}
\exp \left[-\frac{150}{z_h^2} \left( z - \tfrac{1}{3} z_h \right)^{2}  \right]\,,
\ee
where $z_{h}=\frac{2^{1/2}}{3^{1/4}}{\cal E}^{1/4}$. As $B$ is supported at intermediate  values of $z$, naive intuition from the physics of linear wave equations would suggest that the wave packet splits into two: one propagating inwards and the other propagating outwards. The one propagating outwards is expected to eventually reach the boundary, bounce back and fall into the bulk. Both wave packets will be eventually absorbed by the event horizon (which is guaranteed to be present given that ${\cal E} \neq 0$) leading to the increase in its area.

\begin{figure}
\begin{centering}
\includegraphics[width=8cm]{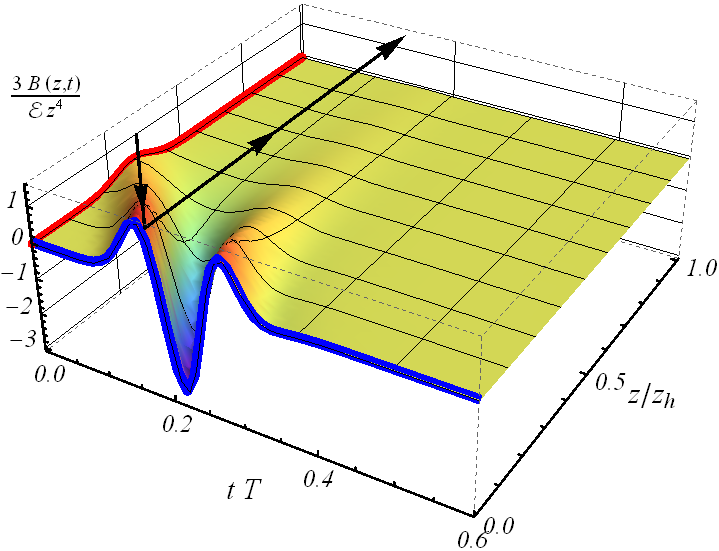}\includegraphics[width=8cm]{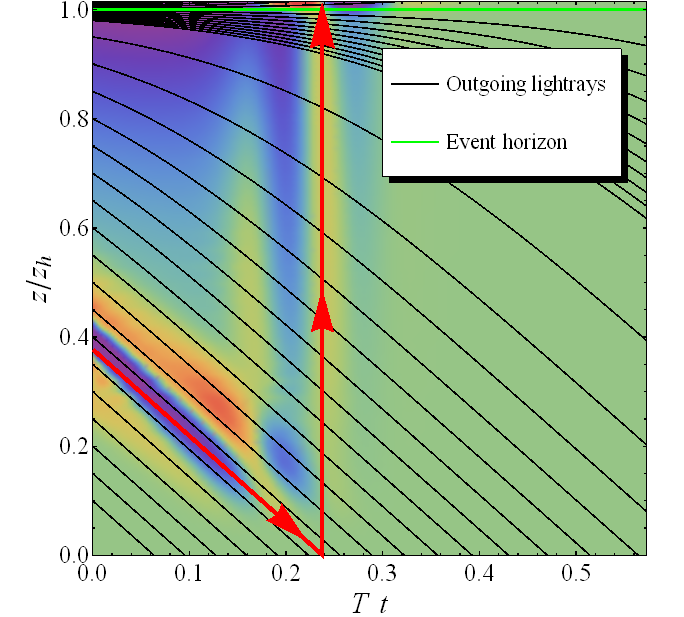}
\par\end{centering}
\caption{The left figure shows $B$ as a function of time and radial coordinate for the initial profile \eqn{sampleprofile}, which is shown as a thick red curve at $t=0$ and which is initially localized near $z = \tfrac{1}{3} z_{h}$. The blue curve at the boundary ($z=0$) depicts the pressure anisotropy as a function of time in the gauge theory. The right figure shows the Kretschmann scalar (with the value for an equilibrium black brane with the same energy density subtracted) as a function of time and radial coordinate for the same initial profile. One clearly sees on this plot the wave bouncing off the boundary and falling into the black brane. In the adopted generalized ingoing Eddington-Finkelstein coordinates this happens instantaneously.
\label{fig:bounce}}
\end{figure}

These expectations are confirmed by the outcome of the  numerical simulation, as illustrated by Fig.~\ref{fig:bounce}, which depicts the bulk anisotropy (left plot) and the square of the Riemann tensor, the Kretschmann scalar (right plot). We can clearly see the rise in the curvature due to the outgoing wave packet as it approaches the boundary of AdS. Closer inspection reveals also the presence of a wave packet  resulting from the bouncing off the boundary of the outgoing packet. This wave packet, due to the null nature of our coordinate frame, propagates towards from the boundary to the horizon along lines of constant Eddington-Finkelstein time. Note also that this signal falls through the black brane event horizon without significant scattering. This feature persisted for other choices  of initial states  and seems to be related to the high degree of symmetry of our problem.

It is interesting to note that the initial ingoing part of the wave packet seems to be mostly  taken care of by the solution of the constraints. Indeed, although $B$ is supported only over some small range of $z$ centered around $z_h/3$, the metric functions $A$ and $\Sigma$ deviate from their vacuum values all the way from this point to the horizon, as required by  causality. In contrast, the curvature outside the outgoing wave packet is very close to the curvature of the static black brane.

These observations suggest that the states which take the longest time to thermalize are those that are initially localized close to the horizon on the initial-time slice. An example is provided by $B(t = 0, z) \sim z^{24}$, whose evolution is shown in Fig.~\ref{Bexampderivs} (right). The reason is that the outgoing wave packet needs to escape the neighborhood of the horizon and travel all the way to the boundary to bounce off and finally fall into the black brane horizon. By localizing the initial profiles close to the horizon, the longest isotropization times that we were able to obtain with our numerics, which used rather moderate grids, were about $1.1/T-1.2 / T$, with $T$ the final equilibrium temperature (see Fig.~\ref{fig:histogram1}). 

\section{Linear approximation for holographic isotropization \label{sec.linearEEQ}}

\subsection{Leading order correction to the pressure anisotropy \label{sec.leadingorder}}

Linearizing Einstein's equations in the setup of holographic isotropization can be \emph{formally} phrased as an expansion in the amplitude of perturbations on top of the AdS-Schwarzschild black brane. We thus write
\bea
\label{deltaASigB}
A(t, z) = \frac{1}{z^2} (1 - z^4) + \alpha \, \delta A^{(1)} (t, z) &+& \mathrm{{\cal O}}(\alpha^2),\nonumber \\
\Sigma(t, z) = \frac{1}{z} + \alpha \, \delta \Sigma^{(1)}(t, z) + \mathrm{{\cal O}}(\alpha^2) \quad \mathrm{and} \quad B(t, z) &=& \alpha \, \delta B^{(1)} (t,z) + \mathrm{{\cal O}}(\alpha^2),
\eea
where $\alpha$ is a formal parameter counting the order in the amplitude expansion.

The smallness of the initial data can be physically quantified by either measuring the total entropy production on the event horizon or by following the amplitude of the pressure anisotropy during the evolution process and comparing it to the energy density. It is important to restress that we want to use the linearized approximation without necessarily restricting to the initial data being small perturbations of the AdS-Schwarzschild black brane, precisely in the spirit of the original close-limit approximation \cite{Price:1994pm,Anninos:1995vf} but now in the context of AdS spacetime.

The initial data for the full nonlinear Einstein's equations are given by specifying the energy density $\ed$ and the form of $B$ as a function of the radial coordinate on the initial-time slice. As anticipated earlier, the main motivation for choosing $B$ over $\Sigma$ in specifying the initial data was that the former appears quadratically in the constraint \eqref{Ct}. This feature persists also with the other components of the Einstein's equations apart from the equation \eqref{Beq}, which immediately leads to
\be
\delta A^{(1)}(t,z) = 0 \quad \mathrm{and} \quad \delta \Sigma^{(1)}(t, z) = 0.
\ee
$\delta B^{(1)}(t,z)$ on the other hand remains nontrivial and is a solution of the equation \eqref{Beq} with $A$ and $\Sigma$ set to their form in the AdS-Schwarzschild background given in \eqref{eq.AdSSchwarzschildComponents}.

The initial condition for this equation is \emph{the same} as the initial condition for the full Einstein's equations, i.e.
\be
\delta B^{(1)}(t = 0, z) = B(t = 0, z).
\ee
The energy density $\ed$, which is constant in our setup and is the remaining part of the initial state specification, is already included in the background we linearize on top of.

In full detail, the equation for $\delta B^{(1)} (t, z)$ reads (with the choice of units $\ed = \frac{3}{4}$)
\be
\label{eq.B1}
\frac{1}{2 z} (3+z^{4}) \, \partial_{z} \delta B^{(1)} - \frac{1}{2} (1-z^{4}) \, \partial_{z}^{2} \delta B^{(1)} - \frac{3}{2 z} \, \partial_{t} \delta B^{(1)} + \partial_{z} \partial_{t} \,\delta B^{(1)} = 0 \,.
\ee
This is a simple, first order in time, partial differential equation which can be solved once supplemented with the Dirichlet boundary condition at $z = 0$. To improve the stability in our computations we actually worked with
\be
\delta B_{reg}^{(1)}(t, z) = \tfrac{1}{z^3} \delta B^{(1)}(t, z)
\ee
and required that $\delta B_{reg}^{(1)} (t, z = 0) = 0$. The other boundary of the computational grid is put inside the event horizon of the background solution.

\begin{figure}
\begin{centering}
\includegraphics[width=15cm]{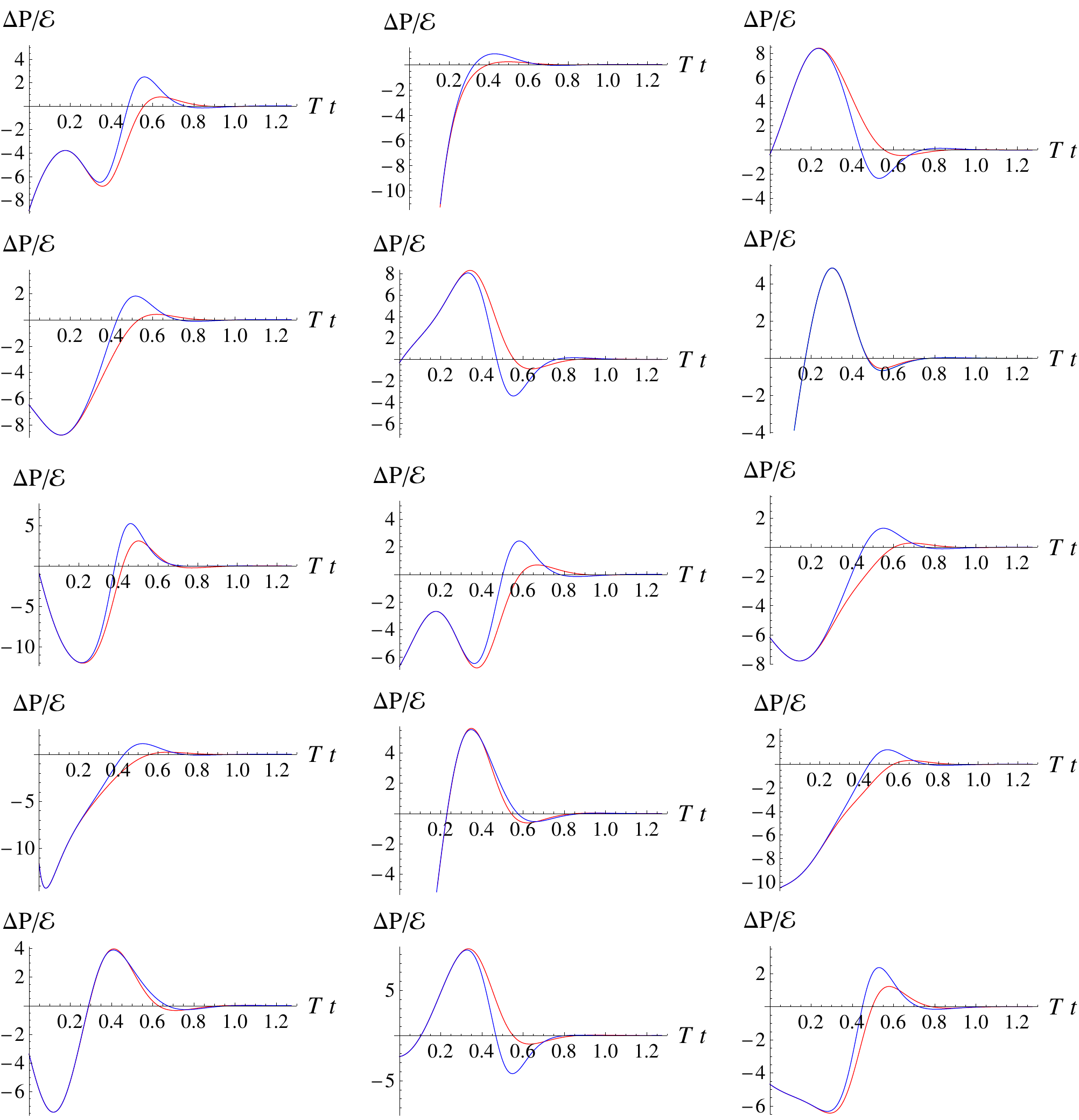}
\par\end{centering}
\caption{Comparison between the time evolution  of the pressure anisotropy predicted by the linear equation \eqref{eq.B1} (red) and the full result (blue) for 15 different initial conditions. The leading order linearized Einstein's equations predict both qualitative and quantitative features of the dynamics of the dual stress tensor in our setup. A more thorough scan of the initial conditions (as shown in  Fig.~\ref{fig:histogram1}) did not reveal any instances in which the linearized approximation broke down.
\label{fig:15profiles}}
\end{figure}

\begin{figure}
\begin{centering}
\includegraphics[width=13cm]{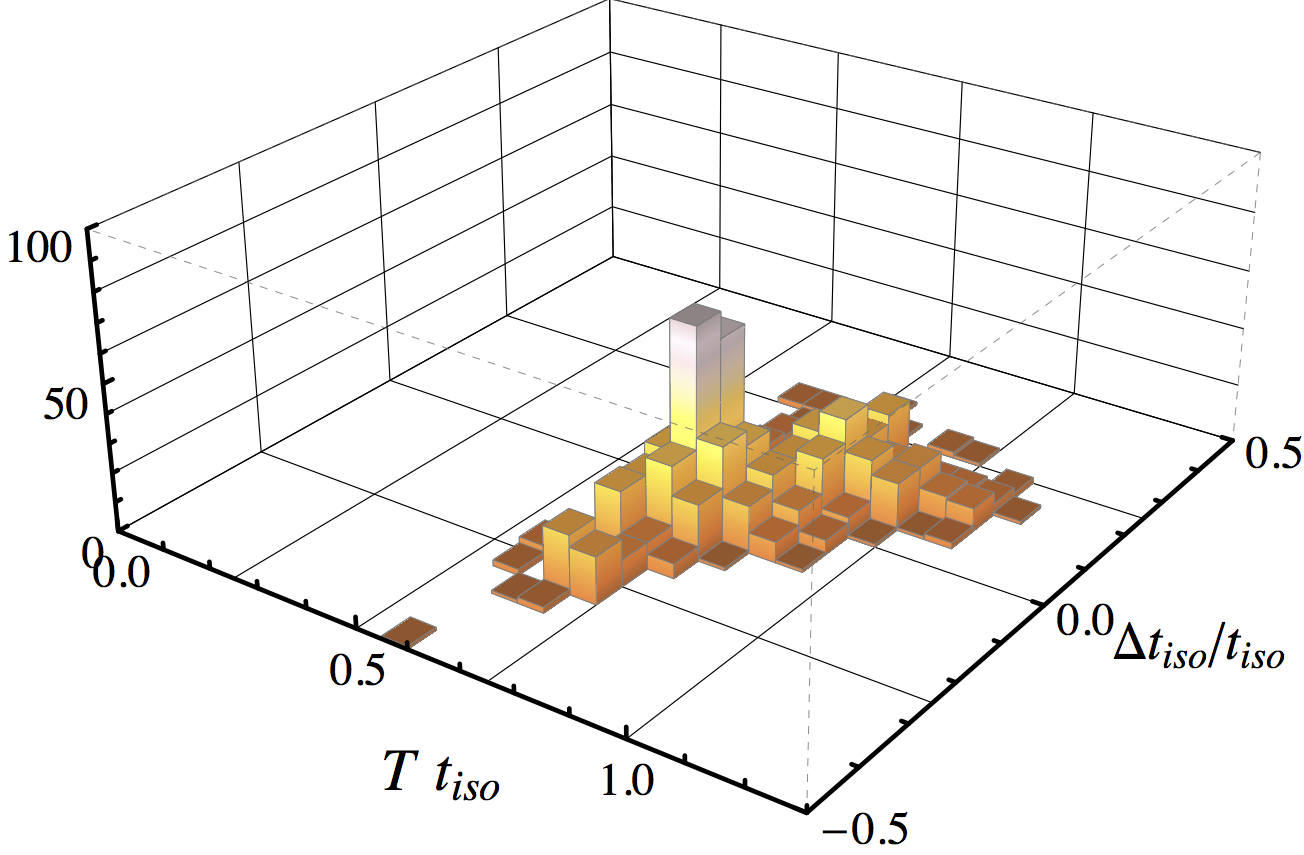}\newline
\includegraphics[width=13cm]{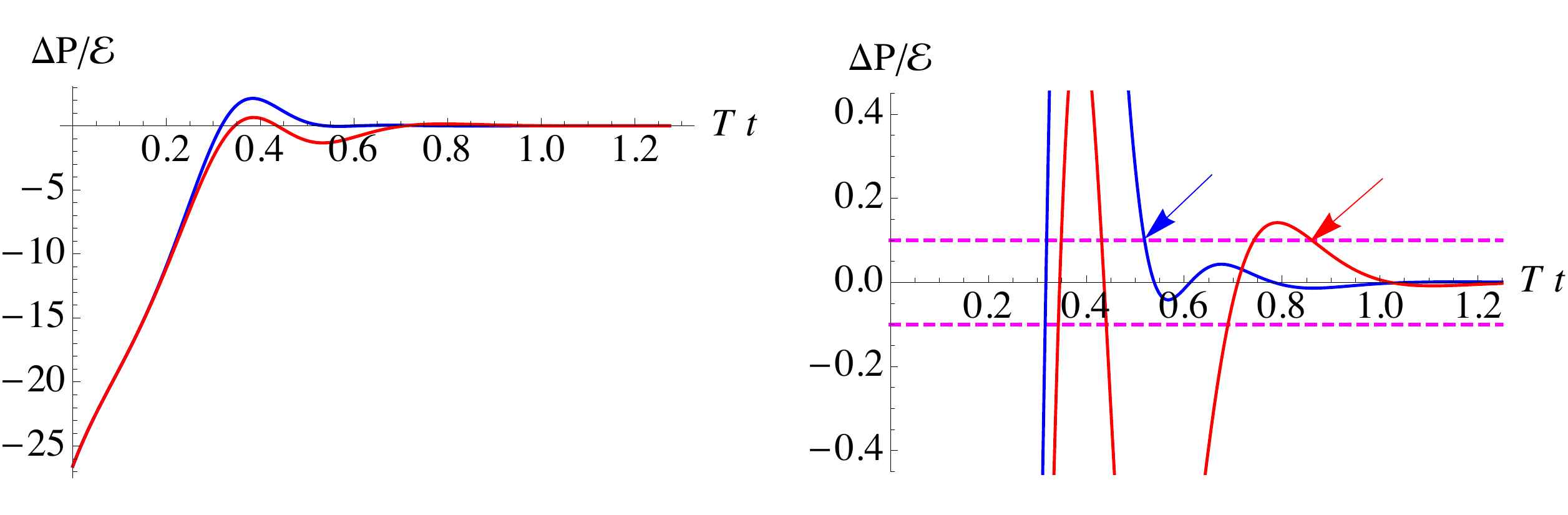}
\par\end{centering}
\caption{(Top) $\Delta t_{iso}$ is the difference between the isotropization time predicted by the full and the linear equations. The height of each bar in the histogram indicates the number of initial states for which the evolution yielded values in the corresponding bin. The total number of initial states is more than 800. We see both that holographic isotropization proceeds quickly, at most over a time scale set by the inverse temperature, and that the linearized Einstein's equations correctly reproduce the isotropization time with a 20\% accuracy in most cases. Note that the histogram is based on a different sample of  initial states than those originally considered in \cite{Heller:2012km}. In particular, we incorporated the binary search algorithm absent in \cite{Heller:2012km} and were  stricter about the maximum violation of the constraint that we allowed.
\newline
(Botom) Close inspection of one of the few profiles for which the linearized approximation seemingly fails by more than 20\% ($\Delta t_\mt{iso}/t_\mt{iso} = -0.5$) shows that it is the imperfect isotropization criterium which leads to the mismatch rather than the failure of the linear approximation. Indeed, the left plot shows that, on the scale of the initial anisotropy, the linear result yields a good approximation. However, the isotropization criterium makes no reference to this scale, and results in a 50\% difference in the isotropization times,  indicated  by the arrows on the right plot. See  \cite{Heller:2012je} for a related discussion of subtleties involved in defining the thermalization (or more accurately hydrodynamization) time in a similar setup.
\label{fig:histogram1}}
\end{figure}

Equation \eqref{eq.B1} captures the leading order dynamics of the whole stress tensor, as the stress tensor is completely specified in terms of the energy density $\ed$ (encoded in the background) and the pressure anisotropy $\deltap (t)$ (encoded in $\delta B^{(1)}$).

In our previous work \cite{Heller:2012km} we compared the predictions for $\deltap(t)$ following from the nonlinear Einstein's equations with the ones obtained using \eqref{eq.B1} and we found that the results agreed with a $20\%$ accuracy for the \emph{vast majority} of the profiles we considered. The match of the qualitative features can be seen in Fig.~\ref{fig:15profiles}. The histogram in 
Fig.~\ref{fig:histogram1} shows the quantitative match for the isotropization time for more than 800 non-equilibrium initial states which are all large perturbations of the AdS-Schwarzschild black brane. We see that the linearized Einstein's equations predict the isotropization time with a $20 \%$ accuracy in the vast majority of cases.

An interesting feature visible in Fig.~\ref{fig:15profiles}, as well as earlier in Fig.~\ref{fig:bounce}, is that the pressure anisotropy obtained from the linearized Einstein's equations \emph{always} follows closely the full result at early times and, if at all, the curves start to differ only at some transient time. We understand this feature as a natural consequence of the fact that, due to the fixed asymptotics, the dynamics is approximately linear near the boundary of AdS, which is the bulk region causally responsible for the early-times dynamics. The pressure anisotropy curves start to differ only after the boundary is reached by a signal propagating from the interior of the geometry, as seen in 
Fig.~\ref{fig:bounce}, and this is the moment at which in the full analysis nonlinear effects become most visible. Our main result can be phrased as the surprising statement that these linearities did not result in a large effect on the boundary stress tensor for any of the initial states that we analyzed.

\subsection{Connection with quasinormal modes}

\begin{figure}[h]
\begin{centering}
\includegraphics[width=7.0cm]{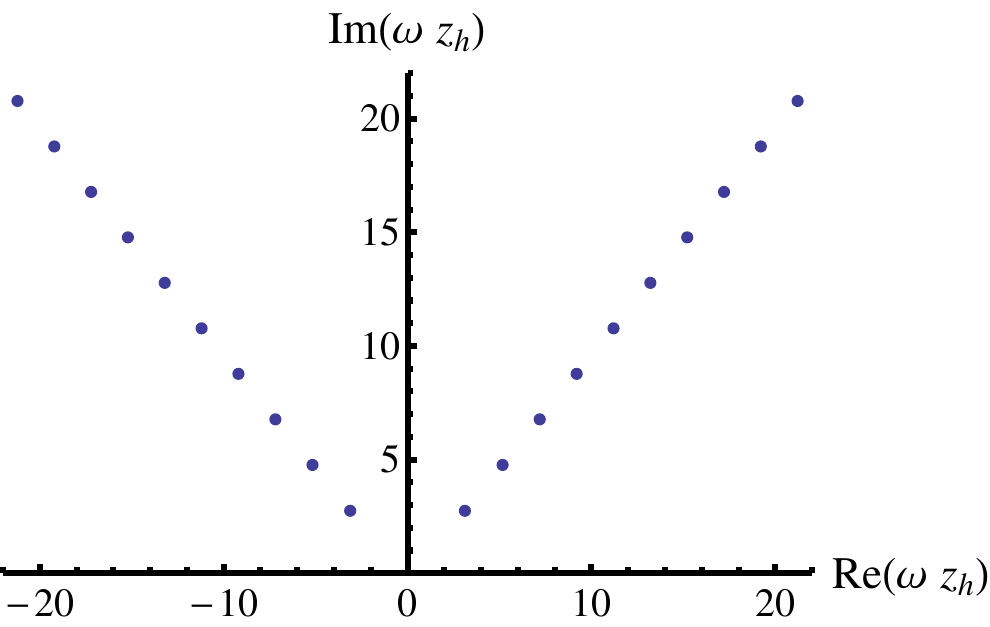}\includegraphics[width=8.0cm]{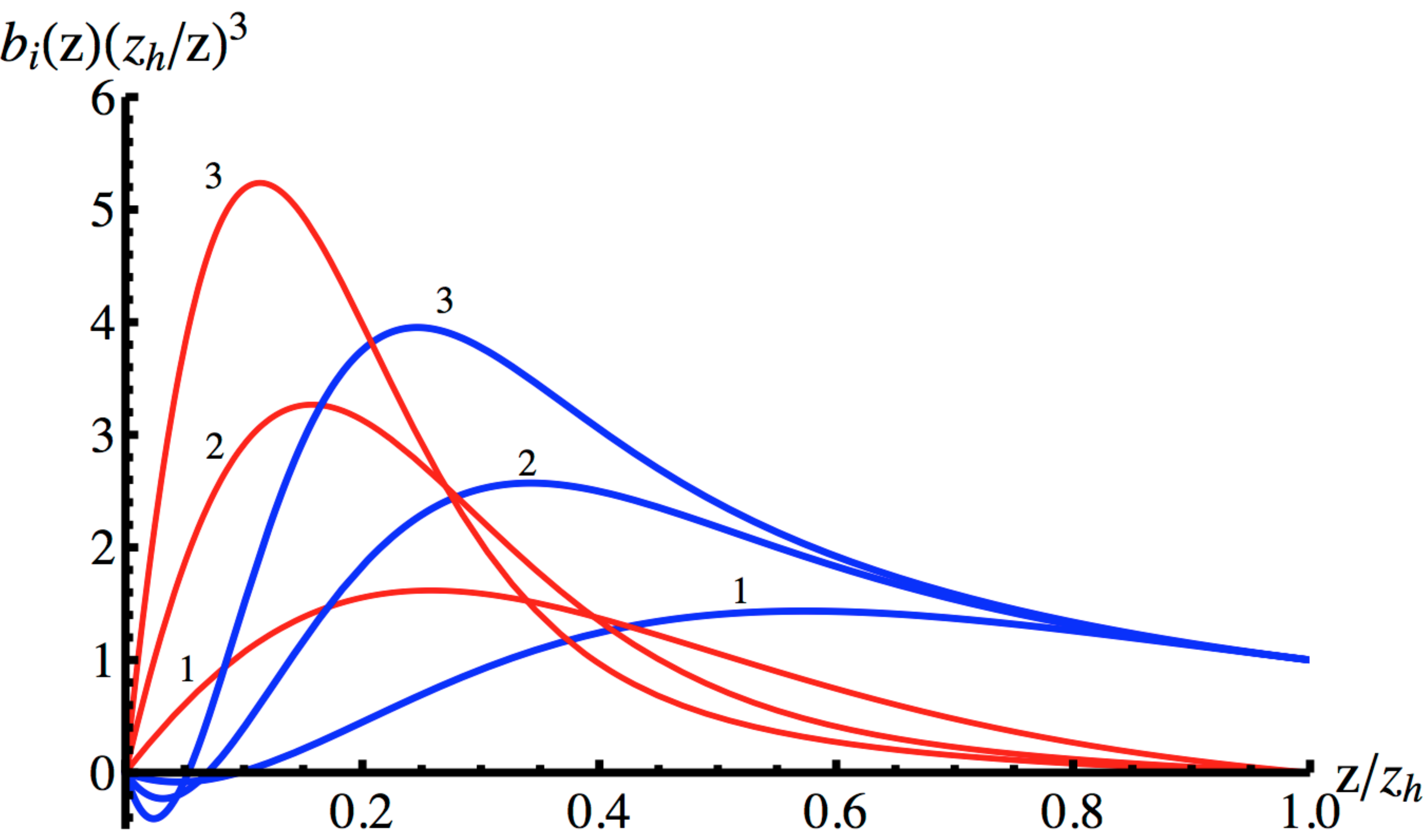}
\par\end{centering}
\caption{The plot on the left shows the frequencies of the ten lowest quasinormal modes including their complex conjugates. The mode with the smallest negative imaginary part will by the dominant mode at late times. Notice that the spacing between the modes is approximately constant (it differs by about 0.1\%). The plot on the right displays the lowest three quasinormal modes as a function of the radial coordinate $z$, where blue and red denote their real and imaginary parts. The normalization we use is such that the real part at the horizon ($z/z_{h} = 1$) is equal to unity, whereas the imaginary part vanishes there. One clearly sees that higher modes (which decay faster) are more dominant near the boundary.}\label{fig:lowestQNMs}
\end{figure}

Equation \eqref{eq.B1} can be solved either as an evolution equation given some initial profile for $\delta B^{(1)}$, as discussed in the previous section, or by decomposing $\delta B^{(1)}$ as a superposition of modes with factorized time dependence:
\be
\label{eq.qnmansatz}
\delta B^{(1)}(t,z) \sim e^{i \omega_{j} t} \, b_{j}(z).
\ee
These modes are known as quasinormal modes, and they are characterized by the requirements that they be normalizable near the  boundary ($z = 0$) and that they obey ingoing boundary conditions at the event horizon ($z = 1$).\footnote{In the ingoing Eddington-Finkelstein coordinates the ingoing condition at the horizon is equivalent to regularity of the solution at the horizon \cite{Hartnoll:2009sz}.}  The latter condition makes the frequencies $\omega_{j}$ complex with imaginary parts responsible for the exponential decay in time. The quasinormal modes \eqref{eq.qnmansatz} appear in pairs, as taking the complex conjugate of the equation \eqref{eq.B1} for the quasinormal mode with frequency $\omega_{j}$ leads to the equation for the quasinormal mode with frequency $-\omega_{j}^{*}$. This feature can be seen in Fig.~\ref{fig:lowestQNMs}.

In the context of gravitational collapse, the \emph{lowest} quasinormal modes are known to govern the late-time decay of black hole perturbations (see e.g. \cite{lrr-1999-2}) and this is also expected in the current setup. On the other hand, the results from \cite{Heller:2012km}, reviewed in the previous section, suggest that the equation \eqref{eq.B1} predicts rather well the full time dependence of the large-$z$ behavior of the warp factor $B$. Hence a natural question is what is the  quasinormal mode content of the perturbations that we considered.

In order to answer this question we followed the prescription of \cite{Horowitz:1999jd} and computed the lowest 10 quasinormal modes \eqref{eq.qnmansatz} by solving equation \eqref{eq.B1} for the ansatz \eqref{eq.qnmansatz} in the near-horizon expansion and evaluating the resulting expression at the boundary to find $\omega_{j}$'s leading to normalizable modes. The (somewhat arbitrary) normalization of our modes is fixed by demanding that at the horizon $(z = 1)$
\be
\label{eq.qnmnormalization}
b_{j}(1) = 1.
\ee
On Fig.~\ref{fig:lowestQNMs} we plot the obtained frequencies $\omega_{j}$ of the lowest 10 quasinormal modes, as well as bulk profiles for the real and imaginary parts of $b_{1}(z)$, $b_{2}(z)$ and $b_{3}(z)$ normalized according to \eqref{eq.qnmnormalization}.

The idea now is to use the quasinormal modes to decompose  solutions of \eqref{eq.B1}, i.e.~to write a solution of \eqref{eq.B1} in the form
\be
\label{eq.qnmexpansion}
\delta B^{(1)}_\mt{QNM}(t, z)=\text{Re}\left[\underset{i=1}{\overset{N_\mt{QNM}}{\sum}}c_{i}\, b_{i}(z)\, e^{i\omega_{i}t}\right],
\ee
where we truncated the expansion at some $N_\mt{QNM}$, although formally we could set \mbox{$N_\mt{QNM} = \infty$}. In our calculations we used $N_\mt{QNM} = 10$. 

One can view \eqref{eq.qnmexpansion} as a further simplification as compared with solving numerically \eqref{eq.B1}, which approximates very well (within $20\%$) the full Einstein's equations when it comes to predicting the form of the dual stress tensor. The reason for this extra simplification  is that now the solution is specified by providing a few complex numbers\footnote{One may construct exceptional initial profiles, which are for instance very close to the boundary, or very rapidly oscillating. Including more quasinormal modes (taking $N_\mt{QNM}$ in \eqref{eq.qnmexpansion} somewhat bigger than $10$) would allow us to treat these cases more accurately.} (say 10 complex coefficients $c_{j}$'s) which due to the linearity of the problem can be fitted on the initial-time slice to $B(t=0,z)$.

As a way of generating coefficients $c_{j}$'s we minimized
\be
\label{eq.leastsquaresfit}
\int_{0}^{1} \frac{dz}{z^3} \left| B(t=0, z) - \delta B^{(1)}_\mt{QNM}(t = 0, z) \right| 
\ee
by using the least squares method on a discrete sample of the radial position $z$. Naturally, one needs far more points
than the number of quasinormal modes included in \eqref{eq.qnmexpansion}. 

The subtlety in using \eqref{eq.leastsquaresfit} lies in the choice of the mutiplicative factor under the integral, which we set to be $1/z^{3}$. We checked that both $1/z$ and $1/z^4$ do not work well, as the first one does not take sufficiently into account and the other overcounts the near-boundary behavior of $B(t = 0, z)$. On the other hand, $1/z^2$ seems to work equally well as $1/z^{3}$, but for definiteness we focused here on the latter.

\begin{figure}[h]
\begin{centering}
\includegraphics[width=5cm]{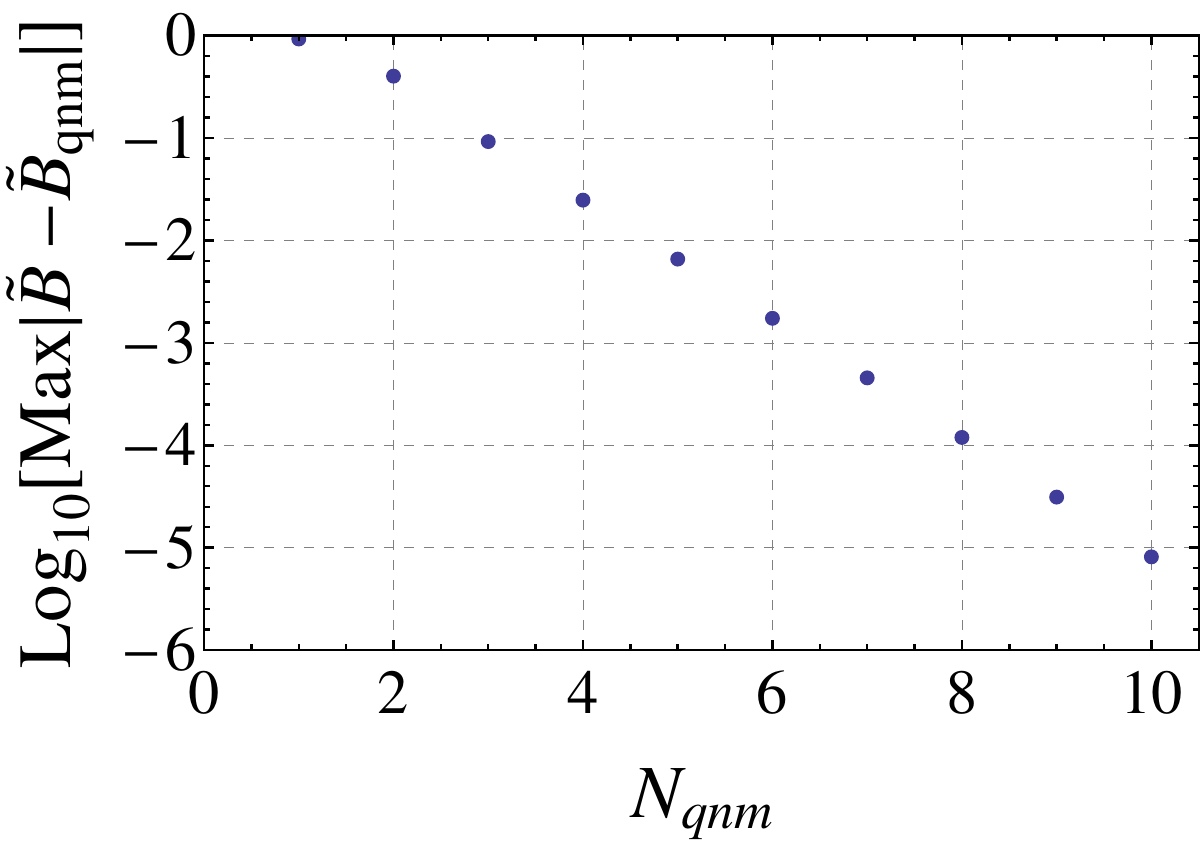}\,\includegraphics[width=5cm]{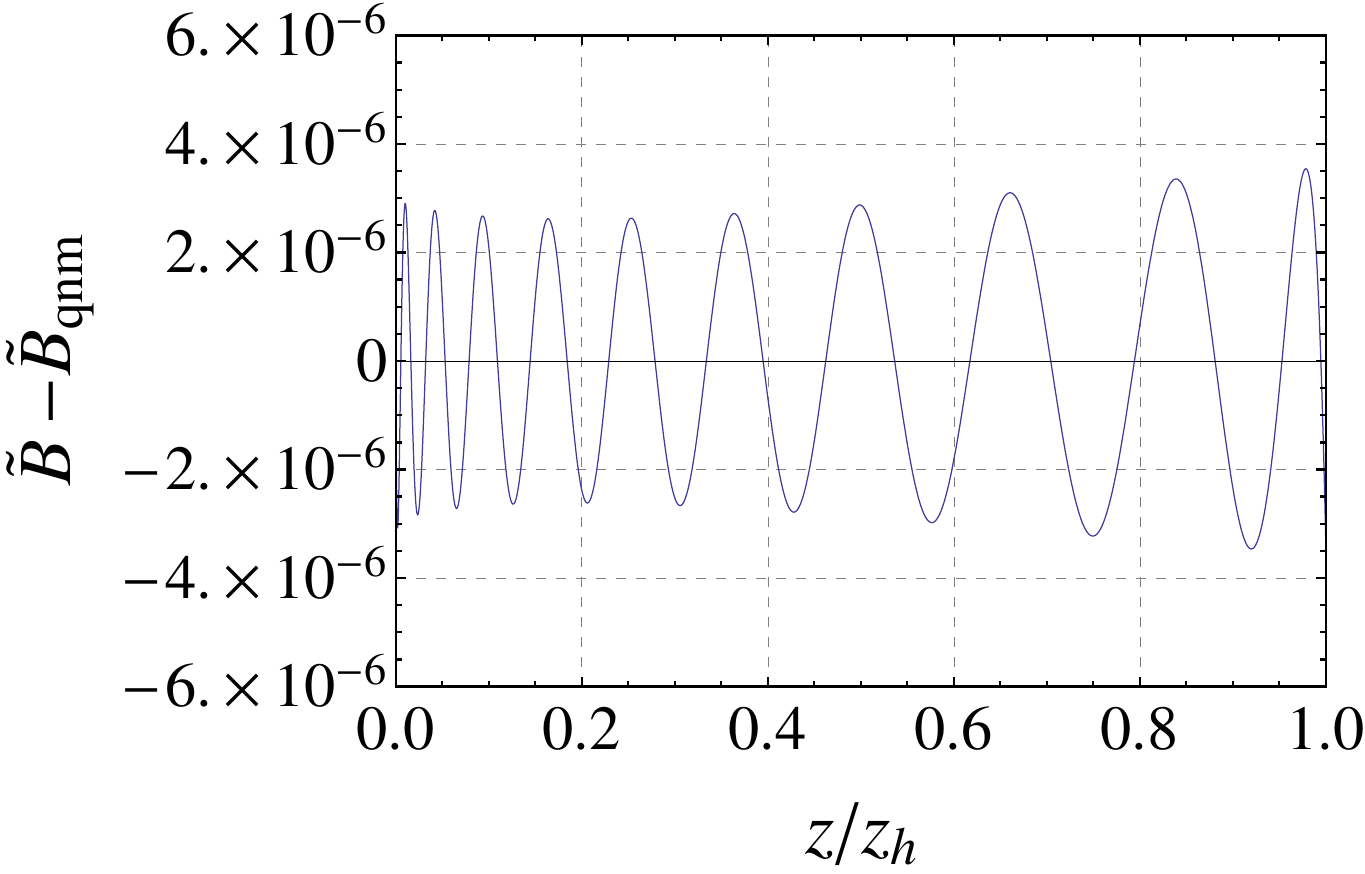}\includegraphics[width=5cm]{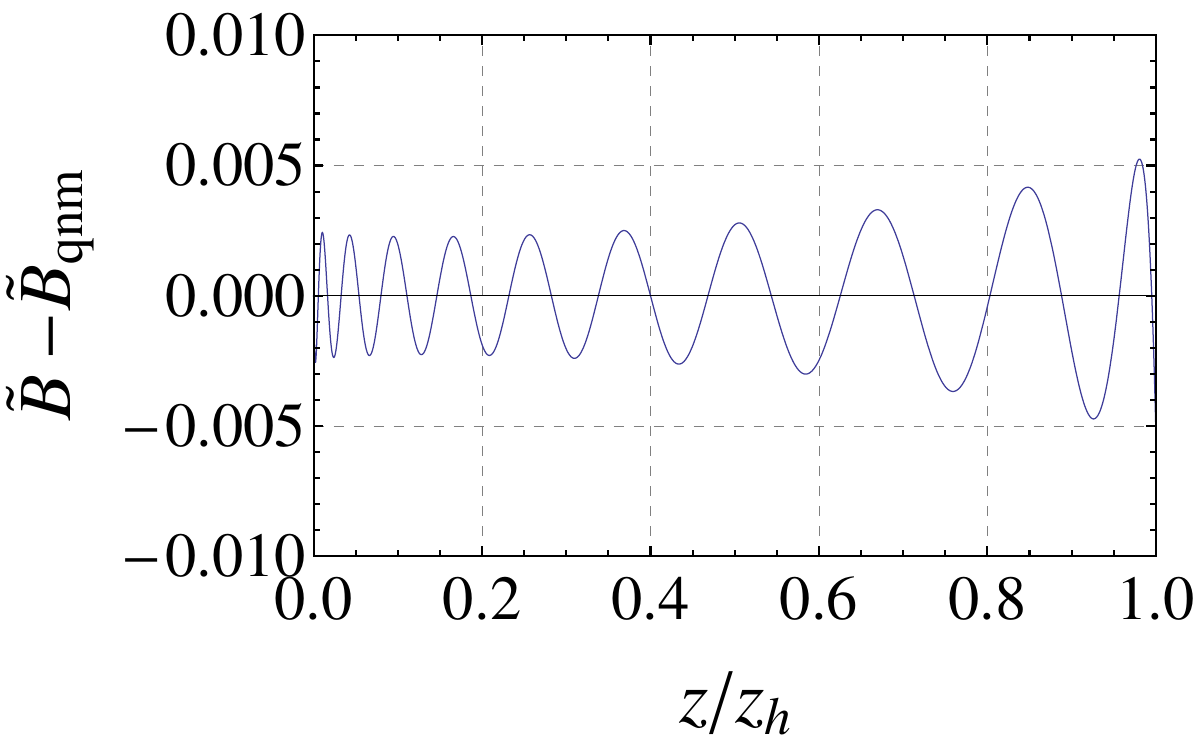}
\par\end{centering}
\caption{The plot on the left displays the maximum of the error when approximating
 $\tilde{B}(z)=B(z)/z^{3}$ by the first $N_\mt{QNM}$ (complex) quasinormal modes, with $B(t = 0, z)=-2 a_4 z^{4}$. The plot in the middle  shows the error for the same profile as a function of the bulk coordinate $z$ while using the 10 lowest quasinormal modes. The right plot displays the error for $B(t=0, z)=z^{25}$ and shows clearly that a profile which is dominated in the IR is much harder to fit by the quasinormal modes. This causes oscillations in the evolution, as can be seen in Fig.~\ref{fig:examples}.
\label{fig:QNMerror}}
\end{figure}

Fig.~\ref{fig:QNMerror} displays the difference between $B(t = 0, z)$ and $\delta B^{(1)}_\mt{QNM}(t = 0, z)$ as a function of  the number of  quasinormal modes in two representative examples. Clearly, if a good fit is possible, then the profile \eqref{eq.qnmexpansion} will solve the linearized Einstein's equations nicely since each quasinormal mode solves them individually. 

\begin{figure}[h]
\begin{centering}
\includegraphics[width=7.5cm]{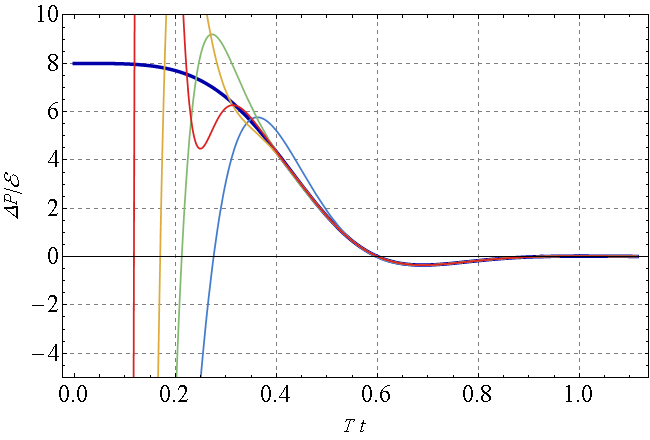}\includegraphics[width=7.5cm]{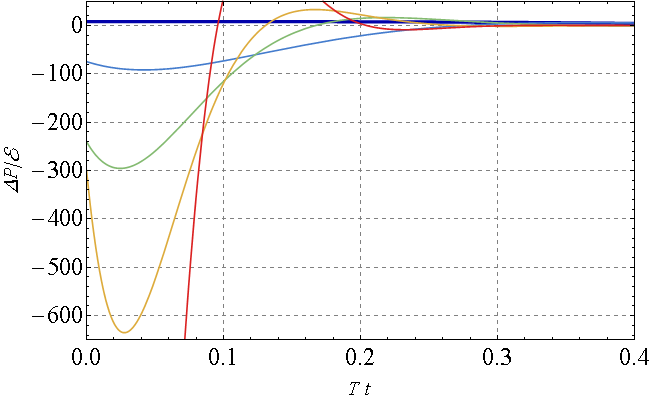}
\par\end{centering}
\caption{On the left one sees the pressure anisotropy $\Delta{\cal P}/{\cal E}$ as predicted by
the linearized evolution, or indistinguishably by the sum of the lowest
10 quasinormal modes as a thick blue line. One can also see there the sum
of the first 1 (blue), 2 (green), 3 (orange) or 4 (red) quasinormal
modes. As becomes apparent, the late time dynamics is well approximated already by keeping only the lowest quasinormal mode, but if one uses more the fit starts matching earlier. Note that the coefficients are computed such that the sum of the 10 fits the initial state best. \newline 
On the
right we plot the individual quasinormal modes with the same coloring.
One clearly sees that each of them carries very large anisotropy,
but that their interference matches the linearized solution. 
\label{fig:QNMindividual}}
\end{figure}

\begin{figure}[h]
\begin{centering}
\includegraphics[width=15cm]{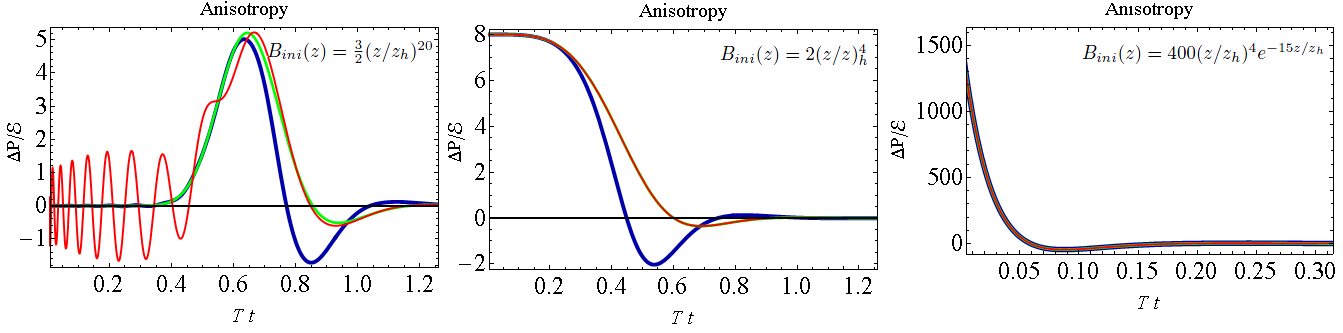}\\
\includegraphics[width=15cm]{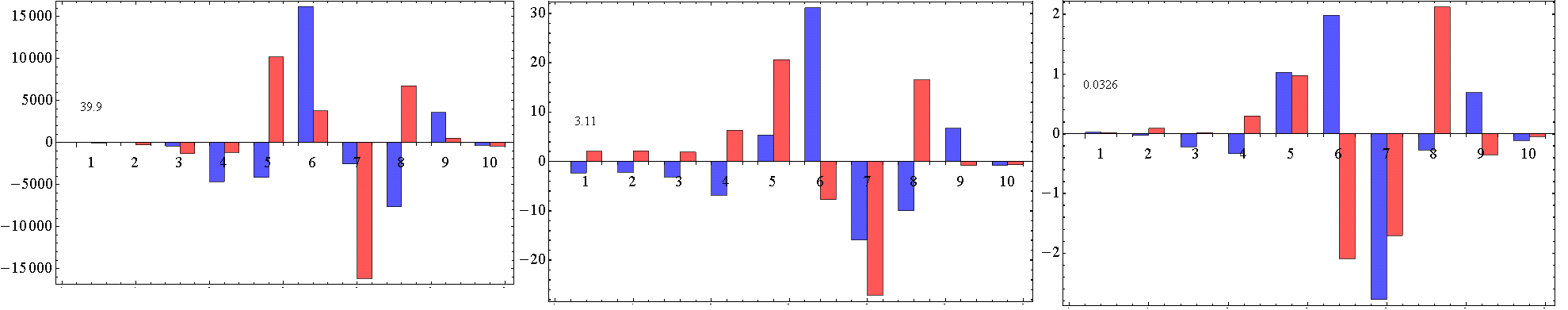}
\par\end{centering}
\caption{In this figure we illustrate anisotropies of the full (blue), linearized (green) and quasinormal mode (red) evolution of three representative initial profiles, located in the IR, spread-out and located in the UV respectively. Clearly, the initial profile located in the IR takes some time before exciting the anisotropy at the boundary, which also explains the late thermalization. The UV profile can have a very large anisotropy, but isotropises very fast. \newline
These features are nicely described when looking at the quasinormal modes coefficients $c_j$'s (below, real (blue) and imaginary (red) part). For the IR profile each individual contribution is very large, but they interfere in such a way to give only moderate anisotropies. In this way it is also possible to reach isotropization as late as 6 times the lowest QNM e-folding time. We also see that one would need to compute more quasinormal modes to accurately fit this profile. \label{fig:examples}}
\end{figure}

In Fig.~\ref{fig:QNMindividual} we compare the linearized evolution obtained from a direct solution of \eqref{eq.B1} and from a solution based on a decomposition into quasinormal modes. One can see  that the contribution from each individual quasinormal mode can be large, %\footnote{Note that this depends on the normalization of the individual modes.} 
but that the final sum approximates the linearized evolution very well. Finally, in Fig.~\ref{fig:examples} we plot three representative examples, where the profile with $B(t=0,z)$ having support mostly in the IR displays this interference phenomenon particularly nicely.

\subsection{Full leading order correction to the bulk}
As reported in \cite{Heller:2012km} and briefly reviewed in section \ref{sec.leadingorder}, linearizing the Einstein's equations leads to a very significant simplification in our setup when it comes to predicting the dynamics of the dual stress tensor, which is the only excited local operator. A natural question arising is whether also the dynamics of nonlocal observables, such as Wilson loops or the entanglement entropy, can be determined in a satisfactory way by using the linearized Einstein's equations alone. As nonlocal observables probing sufficiently large domain in a dual field theory obtain \emph{direct} contributions from the interior of bulk spacetime, this question translates to which extend the linearized gravity reproduces the full bulk.

Already at the superficial level it is easy to see that in order to answer this question we need to extend the analysis from \cite{Heller:2012km} and include quadratic corrections to linearized Einstein's equations. A simple way to understand it is to notice that at the linear level both $A$ and $\Sigma$ retain their background form and hence the equation \eqref{eq.B1} does not capture the phenomenon of the entropy production. The latter appears only when $A$ or $\Sigma$ are different from their background values, as follows from  equations \eqref{eq.horloc} and \eqref{eq.entropy}.

The lack of entropy production at first order in the amplitude expansion does not matter for initial states for which the initial condition $B(t = 0, z)$ is tiny. In this case the quadratic corrections will be even tinier and as negligible as the actual entropy production in this particular process. However, we already saw that the linearized Einstein's equations do a good job in predicting $\deltap (t)$ also in the cases when the perturbation is not small by any means. In these situations though, they would do a miserable job in predicting the entropy production.

This argument implies that actually the \emph{full} leading order correction to the bulk spacetime in the current setup comprises $\delta B^{(1)}$, which captures the dynamics of $\deltap (t)$ in the leading order, and $\delta A^{(2)}$ and $\delta \Sigma^{(2)}$. The latter quantities are directly responsible for the leading order entropy production in the close-limit approximation in our setup.

From a more general perspective, we will treat the entropy production as an example of a quantity that is highly sensitive to the IR part of the bulk geometry and we expect that the corrections $\delta A^{(2)}$ and $\delta \Sigma^{(2)}$ will be crucial for computing other observables of that sort, including two-point functions, entanglement entropy, Wilson loops, etc.

By looking closer at the full set of nonlinear Einstein's equations \eqref{Eeqns} one can easily anticipate that $\delta \Sigma^{(2)}$ can be obtained by solving the constraint \eqref{Ct} expanded up to the quadratic order in the amplitude
\be
\label{eq.deltaS2}
z \, \partial_{z}^2 \, \delta \Sigma^{(2)} + 2 \, \partial_{z} \, \delta \Sigma^{(2)} = - \frac{1}{2} (\partial_{z} \delta B^{(1)})^{2}.
\ee
This is a second order linear ordinary differential equation in $z$, which, having obtained $\delta B^{(1)}(t,z)$, can be solved using very basic numerical methods on each time slice. The boundary condition for $\delta \Sigma^{(2)}$ is that it vanishes at $z = 0$, as the boundary behavior is always incorporated in the background.

Having both $B^{(1)}(t,z)$ and $\Sigma^{(2)}(t,z)$ we can use  equation \eqref{Aeq} expanded to the quadratic order in amplitude to solve for $\delta A^{(2)}(t,z)$
\bea
\label{eq.deltaA2}
\hspace{-20 pt}z^2 \partial_{z}^{2} \delta A^{(2)} + 2 z \partial_{z} \delta A^{(2)} - 6 \delta A^{(2)} = && - \frac{3}{2} (1- z^{4}) (\partial_{z} \delta B^{(1)})^{2} + 3 \partial_{t} \delta B^{(1)} \partial_{z} \delta B^{(1)} \nonumber \\
&&- \frac{12}{z} (1 - z^{4}) (\delta \Sigma^{(2)} + z \partial_{z} \delta \Sigma^{(2)}) + 12 \partial_{t} \delta \Sigma^{(2)}.
\eea
Again, one obtains a second order, linear, ordinary differential equation in $z$, which can be integrated with the use of basic numerical techniques. The choice of integration constants is dictated by the near-boundary expansion \eqref{nearbdryexpansions}, which at low orders is the same for the linearized and the full Einstein's equations. 

At the second order of the amplitude expansion the correction to $B$ vanishes. The reason is that equation \eqref{Beq} at this order is homogeneous and first order in time, \emph{and} we already used the initial profile for $B$ as the initial condition for $\delta B^{(1)}$ so that $B^{(2)}(t = 0, z) = 0$. Finally, it is worth noting that in order to improve the accuracy of our procedures, in numerical calculations we actually redefined $\delta \Sigma^{(2)}$ and $\delta A^{(2)}$ to be
\be
\label{eq.redefSig2andA2}
\delta \Sigma_{reg}^{(2)} = z^{-5} \delta \Sigma^{(2)} \quad \mathrm{and} \quad \delta A_{reg}^{(2)} = z^{-4} \delta A^{(2)}.
\ee
Although formally we refer to our expansion scheme as `the amplitude expansion', in the spirit of the close limit approximation we are actually interested in evaluating the whole leading order contribution to the geometry ($\delta B^{(1)}$, $\delta \Sigma^{(2)}$ and $\delta A^{(2)}$) for 
 initial data of arbitrary amplitude and in comparing it with the full answer.

\subsection{Leading order correction to entropy production}

Having obtained the geometry at leading order, i.e.~having solved  equations \eqref{eq.B1}, \eqref{eq.deltaS2} and \eqref{eq.deltaA2}, we use \eqref{eq.horloc} and \eqref{eq.entropy} to calculate the position of the event horizon as well as its entropy. 

Contrary to the nonlinear case, we are not guaranteed that the entropy will be a non-decreasing function of time. An easy way to understand it is by noting that the solution of linearized equations corresponds to the solution of full equations with some `matter' stress tensor in the bulk. As this matter is not guaranteed to obey the energy conditions, the approximated entropy may in principle decrease. Moreover, the linearized formula for the area of the event horizon following from 
e.g.~(\ref{eq.entropy}) is not guaranteed to lead to a nonnegative result. 

\begin{figure}
\begin{centering}
\includegraphics[width=15cm]{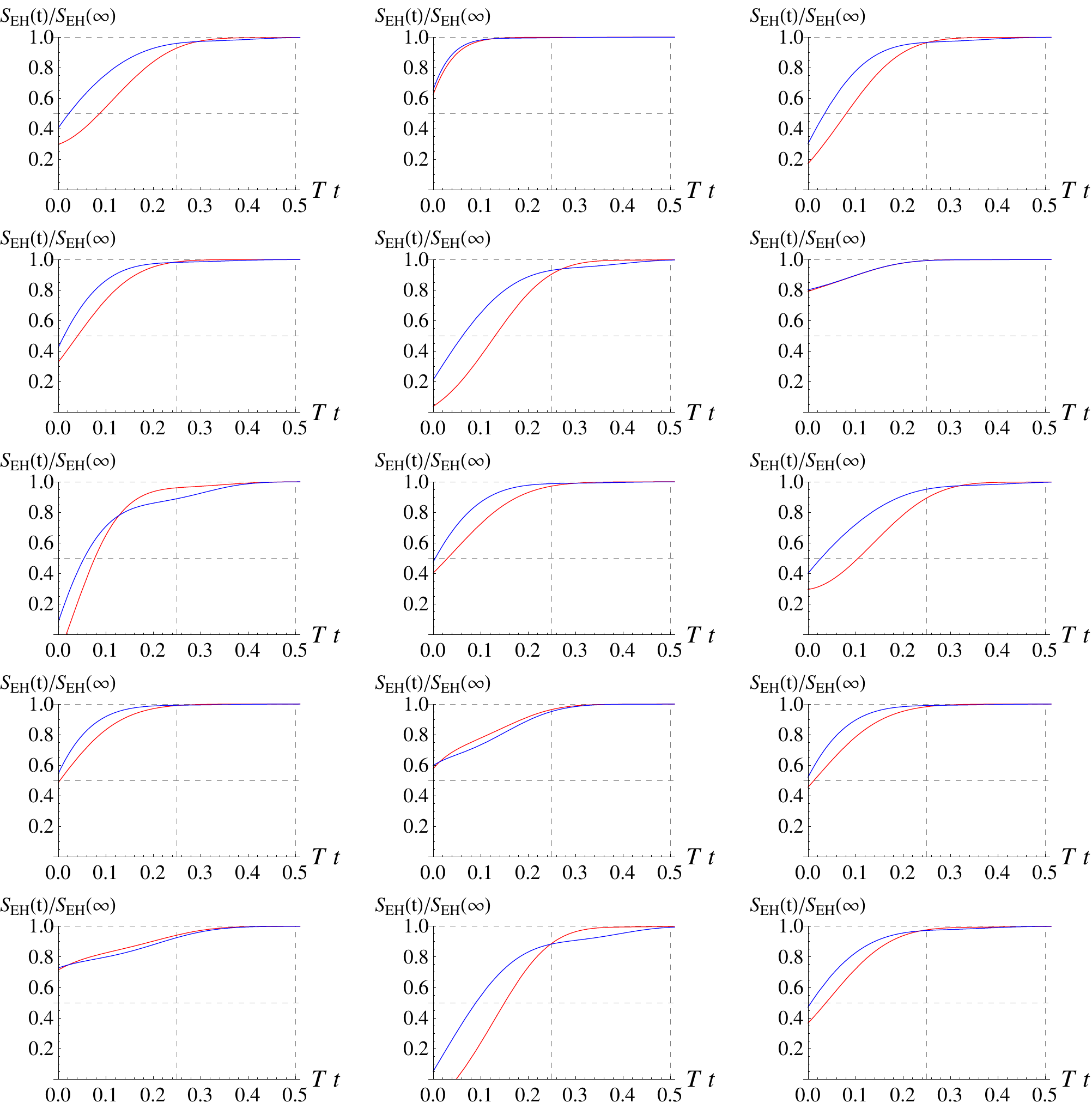}
\par\end{centering}
\caption{Comparison between the leading order prediction for the entropy production coming from the Einstein's equations expanded to quadratic order  (red curves) and the exact entropy production the obtained from the full nonlinear analysis (blue curves), for the same 15 initial states as those in Fig.~\ref{fig:15profiles} with the same ordering. Note that none of the initial conditions can be regarded as a small perturbation either in the sense of the amplitude or the total entropy produced during the equilibration process. On Fig. \ref{fig:histogramAREA} it can be seen that the leading order prediction for the entropy production from linearized Einstein's equations matches remarkably well, within $20\%$, with the full nonlinear result for the vast majority of the initial states that we analyzed.
\label{fig:15profilesAREA}}
\end{figure}

\begin{figure}
\begin{centering}
\includegraphics[width=13cm]{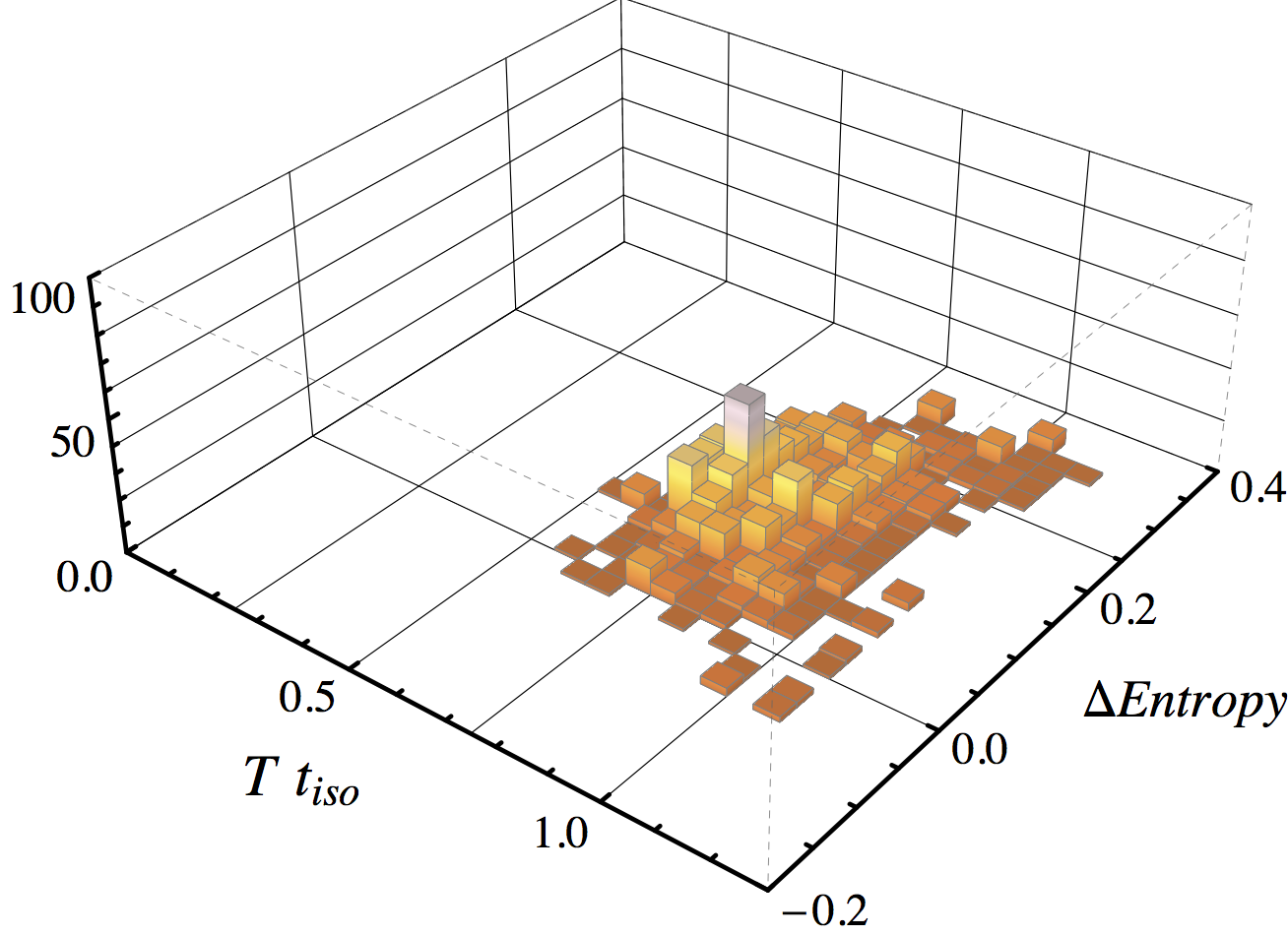}
\par\end{centering}
\caption{Histogram for the error in the entropy production analogous to that in Fig.~\ref{fig:histogram1} for the isotropization time. Both histograms are based on the same sample of more than 800 initial states. We define the error in the entropy production as $\Delta \mathrm{Entropy} = \frac{\Delta s^{(2)} - \Delta s}{\Delta s}$, where 
\mbox{$\Delta s=s(t=\infty) -s(t=0)$} and $\Delta s^{(2)}=s^{(2)}(t=\infty) -s^{(2)}(t=0)$ are the exact and the leading order entropy productions, respectively. We see that the leading order approximation reproduces the correct result with a $20\%$ accuracy for the vast majority of states.
\label{fig:histogramAREA}}
\end{figure}

It turns out, however, that most often these two caveats do not apply, and even if they do, they do not change the results significantly when compared with the full answer, as can be seen in Figs.~\ref{fig:15profilesAREA} and \ref{fig:histogramAREA}, which are the entropy analogues of the anisotropy Figs.~\ref{fig:15profiles} and \ref{fig:histogram1}. The results depicted in Figs.~\ref{fig:15profilesAREA} and \ref{fig:histogramAREA} show that 
the leading order correction consisting of $\delta B^{(1)}$, $\delta \Sigma^{(2)}$ and $\delta A^{(2)}$ reproduces a bulk IR-sensitive observable such as the entropy production with a 
 $20\%$ accuracy for the vast majority of states. The numerical techniques needed for obtaining these leading order approximation are much simpler than needed to obtain the nonlinear results. In particular, the built-in \emph{Mathematica} command {\tt NDSolve} is sufficient. This is already a significant simplification, as the actual procedures used to solve the nonlinear problem required a fully-fledged numerical implementation.

\subsection{Including higher order corrections}

Having computed the full leading order correction, a natural question is whether including higher order terms in the amplitude expansion will get us closer to the full nonlinear result. A direct calculation shows that the first subleading correction to the pressure anisotropy comes from $\delta B^{(3)}$, whereas the first subleading correction to the entropy production comes from the combined effect of including $\delta A^{(4)}$ and $\delta \Sigma^{(4)}$. 

\begin{figure}
\begin{centering}
\includegraphics[width=\textwidth]{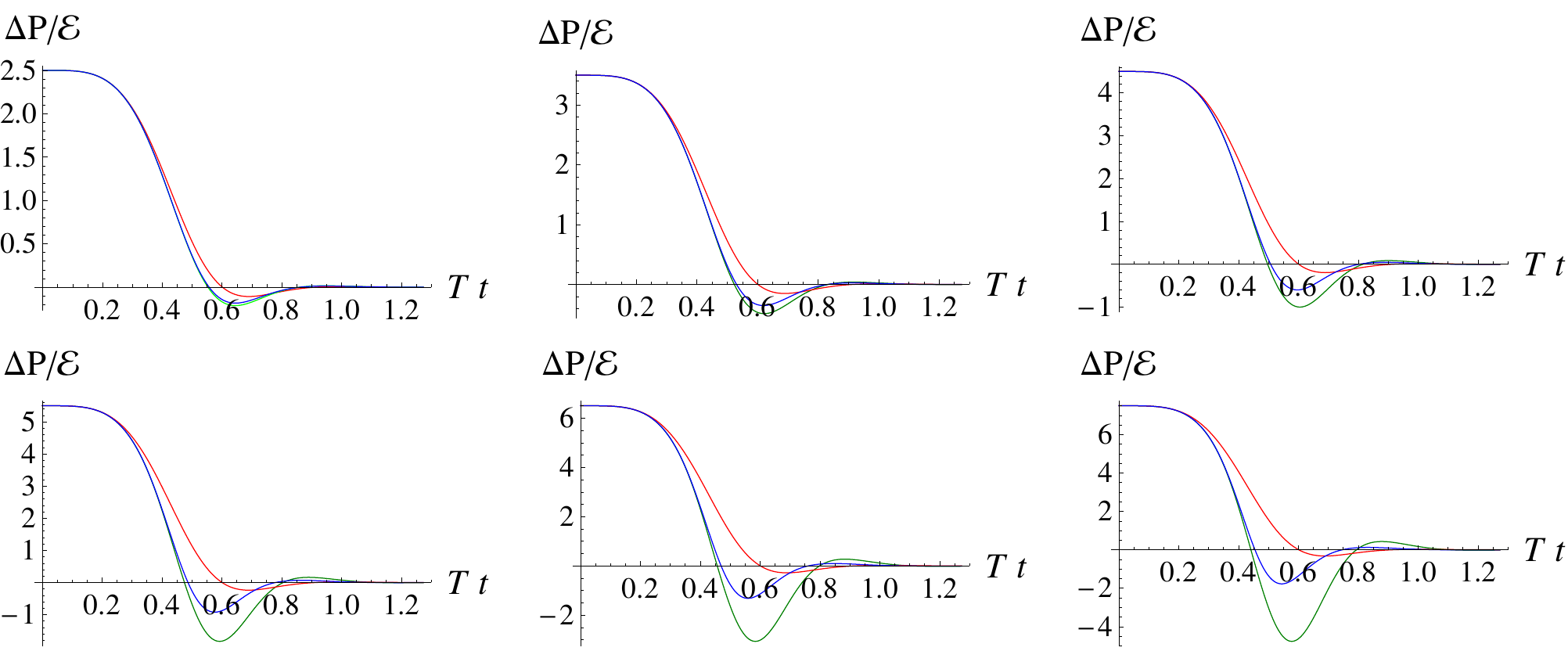}
\par\end{centering}
\caption{Pressure anisotropy as a function of time as predicted by the full Einstein's equations (blue curves), by the first order approximation (red curves) and by the first+third order approximation (green curves). In all cases the initial state is of the form \eqref{eq.Btrial} with an amplitude that increases by $\frac{1}{3}\ed$ from one plot to the next (from left to right and from top to bottom). We see that for small (large) amplitudes adding the third order result improves (worsens) the agreement with the exact result.
\label{fig:incrampDP}}
\end{figure}

\begin{figure}
\begin{centering}
\includegraphics[width=\textwidth]{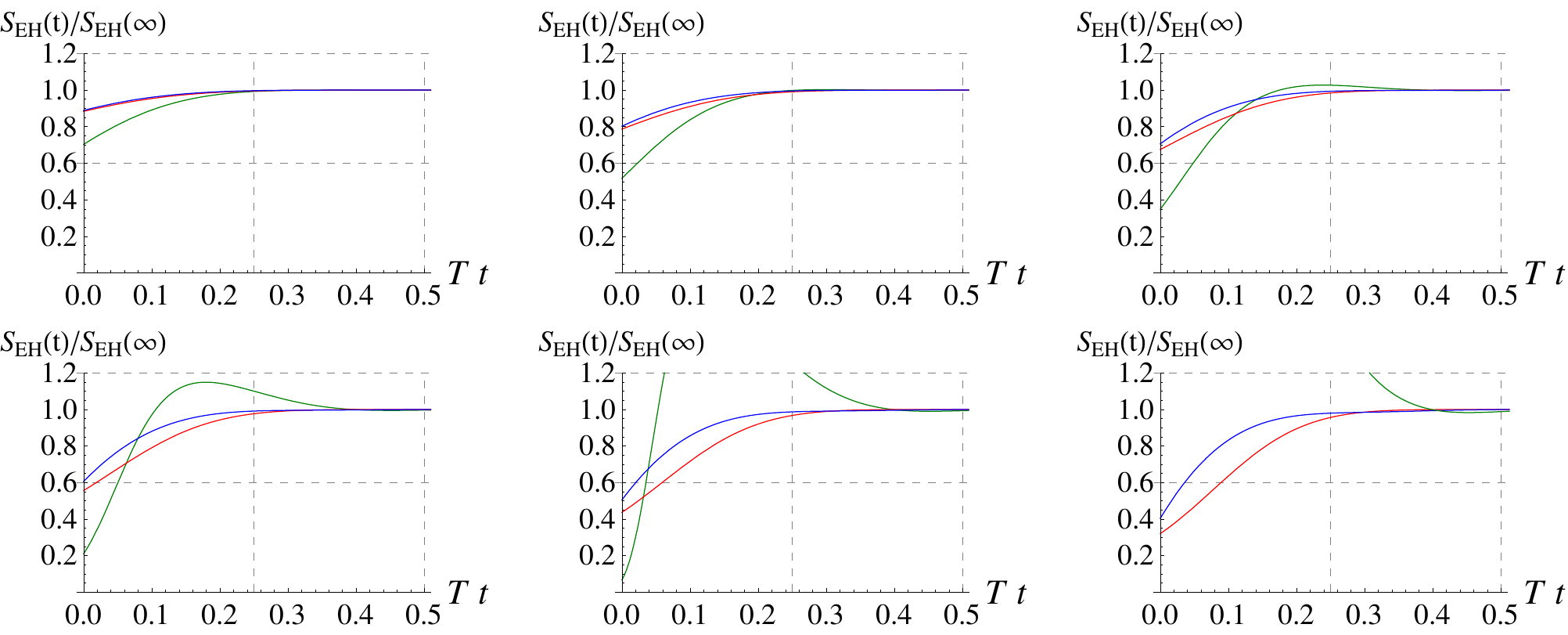}
\par\end{centering}
\caption{Entropy of the event horizon as a function of time as predicted by the full Einstein's equations (blue curves), by the second order approximation (red curves) and by the second+fourth order approximation (green curves). In all cases the initial state is of the form \eqref{eq.Btrial} with an amplitude that increases by $\frac{1}{3}\ed$ from one plot to the next (from left to right and from top to bottom). We see that the second order prediction works better for all cases  shown. 
\label{fig:incrampS}}
\end{figure}

\begin{figure}
\begin{centering}
\includegraphics[width=13cm]{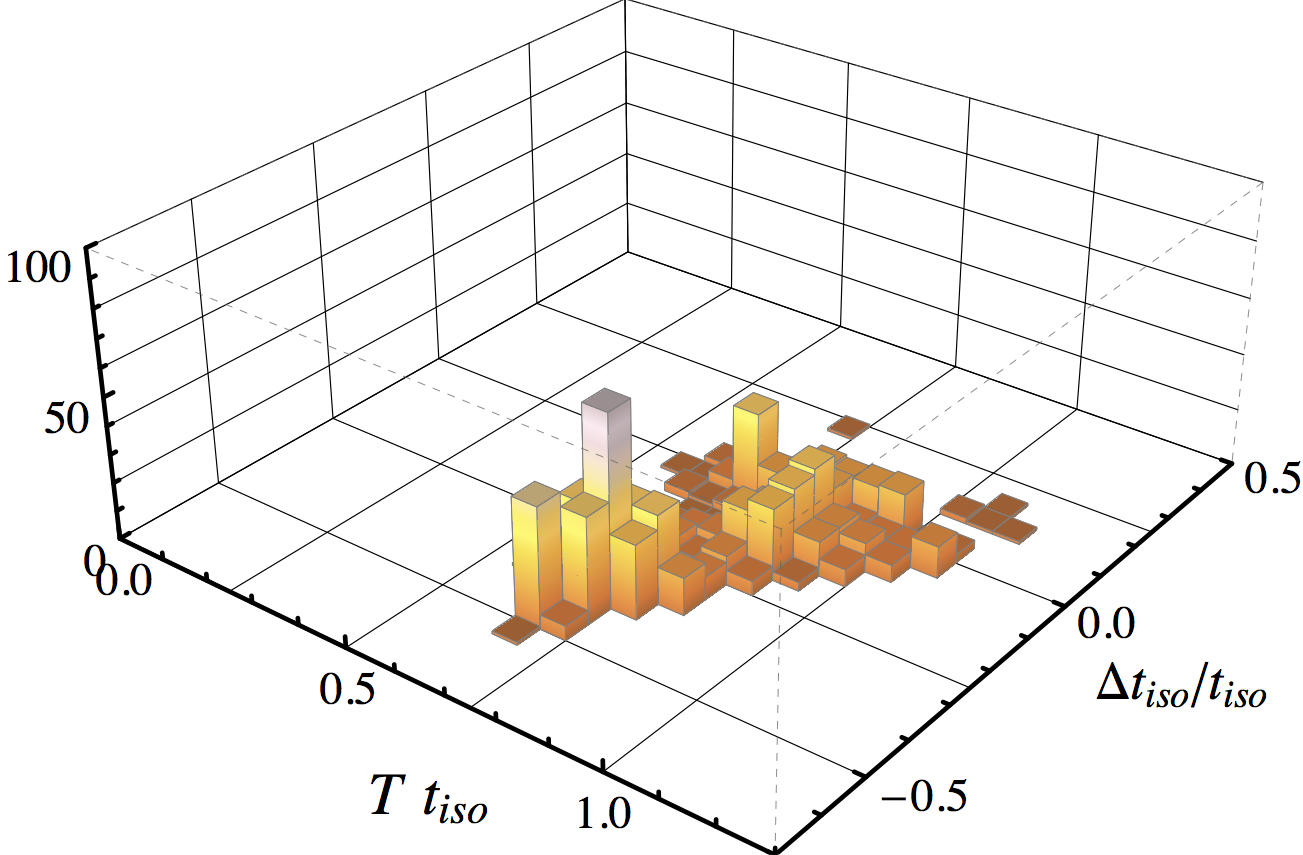}
\par\end{centering}
\caption{Same histogram as in Fig.~\ref{fig:histogram1} except that the first order approximation for the pressure anisotropy is replaced by the first+third order approximation. Comparing with Fig.~\ref{fig:histogram1} we see that the histogram does not become narrower along the $\Delta t_\mt{iso}$ axis after including the third order correction.
\label{fig:histogram_3}}
\end{figure}

In order to understand the role of these subleading corrections, we compared the predictions for the pressure anisotropy from the following  three approaches:
\begin{itemize}
\item nonlinear simulations;
\item leading order solution of linearized equations, 
i.e.~$\delta B^{(1)}$, $\delta \Sigma^{(2)}$ and 
$\delta A^{(2)}$;
\item solution of the linearized equations including the first subleading corrections, i.e.~$\delta B^{(3)}$, $\delta \Sigma^{(4)}$ and $\delta A^{(4)}$\,.
\end{itemize}
We started with some initial $B(t = 0, z)$ and, after each run, we increased its amplitude by fixed amount. For definiteness, in Figs.~\ref{fig:incrampDP} and \ref{fig:incrampS} we show the results for profiles obtained from
\be
\label{eq.Btrial}
B(t = 0, z) = \frac{5}{6} \, {\cal E} z^4
\ee
by increasing the amplitude by $\frac{1}{3} \ed$ at each run. As expected, the leading order approximation predicts well both the pressure anisotropy and  the entropy production for all amplitudes, including the largest ones that  we could simulate. Also unsurprisingly, for small enough amplitudes the subleading correction improves the result as compared to the leading order result. However, for sufficiently large amplitudes the expansion does not seem to converge. In the case of the pressure anisotropy, including the subleading correction overshoots the prediction for the initial states with the largest amplitudes. In the case of the entropy density, including the subleading correction for large-amplitude profiles leads to rather unphysical results since the entropy density decreases over certain periods of time. A further indication that adding subleading corrections  does not improve the result  for generic states comes from Fig.~\ref{fig:histogram_3}, where we see that the histogram for $\Delta t_\mt{iso}/t_\mt{iso}$ does not become narrower along the $\Delta t_\mt{iso}$ axis after including the third order correction for $\deltap(t)$.

With the prospect of studying other more complicated setups, all these results point to the conclusion that it is safe, and generically better, to keep only the leading order terms. 

\section{Summary and open problems}

In this article we presented a thorough study of the simplest example of holographic thermalization process, the holographic isotropization. The primary motivation for these investigations was understanding to which extent the full process of holographic isotropization can be simplified by describing it using the linearized Einstein's equations alone. Such an approximation is an AdS generalization of the closed limit approximation for asymptotically flat spacetimes, in which in certain black hole mergers one linearizes the Einstein's equations on top of the final black hole predicting quite accurately the gravitational wave pattern at infinity. Our previous studies in \cite{Heller:2012km} showed that such generalization of the close limit approximation works remarkably well, i.e.~with a $20\%$ accuracy, when it comes to predicting the time dependence of the expectation value of the boundary stress tensor. 

The main phenomenological motivation for studying holographic thermalization is learning possible lessons about the way the thermalization (or rather hydrodynamization) process proceeds in   relativistic heavy ion collisions at RHIC and LHC. By replacing QCD by a theory with a gravity dual one only expects to obtain either qualitative insights or quantitative ball-park estimates \cite{Mateos:2011bs}. In this sense a $20\%$ accuracy is more than what is needed in order to understand the phenomena we are interested in,  and at the same time may allow to address otherwise technically hard-to-tackle questions. Two examples of such problems in the relativistic heavy ion collisions context are the pre-equilibrium phase of the elliptic flow and the equilibration of transverse-plane inhomogeneities following from event-by-event fluctuations. Solving their holographic analogues in full generality will require complicated simulations of AdS-black hole spacetimes depending on all five bulk coordinates and our hope is that a suitably developed linear approximation may allow us to obtain results with a reasonable accuracy  at a much smaller cost.

In addition to elaborating in detail on our previous studies, the chief aim of the current paper was to establish to what extent our approximation reproduces the full thermalization process, i.e.~not only the behavior of fields in the vicinity of the boundary (responsible for the expectation values of gauge theory operators), but also the entire bulk spacetime (which carries direct information about e.g.~nonlocal probes in the gauge theory). 

In our previous studies we linearized Einstein's equations around  the final-state black brane solution and we kept only first order terms in the amplitude expansion. We then solved these linear equations supplemented with appropriate  AdS boundary conditions and with the same initial conditions as required to solve the full Einstein's equations. In the spirit of the closed-limit approximation the initial conditions were not necessarily small perturbations of the final black brane. This procedure correctly captured the evolution of the expectation value of the gauge theory stress tensor with a 20\% accuracy, but it predicted no entropy production despite the fact that this was often substantial.

From this perspective, our previous studies in \cite{Heller:2012km} provided only a partial contribution to the full leading order approximation to the bulk spacetime, which also includes terms quadratic in the amplitude expansion. These quadratic terms do not contribute to the expectation value of the stress tensor, but they do modify the bulk geometry. In order to probe this effect, in this paper we considered the contribution of the quadratic terms to the entropy production and concluded that this is also reproduced with a 20\%  accuracy.

We also analyzed the effect of higher order corrections in the amplitude expansion. Not surprisingly, we found that for initial states that start off sufficiently close to equilibrium the inclusion of these terms improves the agreement with the nonlinear evolution. However, for large deviations from equilibrium (large amplitudes) the inclusion  of these terms actually spoils the 20\% agreement between the leading order approximation and the exact result. This suggests that in future cases in which a full nonlinear simulation might not be feasible, it might be enough to work with the leading order approximation.

An interesting byproduct of our analysis is that the quasinormal modes, which so far were thought to be responsible only for the late time approach to equilibrium in the expectation values of dual operators, might actually predict quite accurately the full time dependence of the dual stress tensor provided a sufficient number of them is included.

The most important open problem raised by our previous work \cite{Heller:2012km} and the current paper is how to generalize the close-limit approximation to more phenomenologically interesting setups, for example to expanding plasmas such as those considered in \cite{Chesler:2009cy,Heller:2011ju,Chesler:2010bi}. Those setups are qualitatively different form the one we have considered because of two  interrelated features. First, the late time physics will be described by a solution of hydrodynamic equations which is not known a priori in terms of the initial conditions without solving for the evolution of the system.
This contrasts with the current case, in which the final state is a homogeneous thermal state whose energy density is known from the start. Second, the late time state will not be static, so  one does not expect to be able to describe it by linearizing around the AdS-Schwarzschild black brane. Instead, since the energy density of an expanding plasma gets more and more diluted as time progresses, one might naively hope to be able to linearize around the Poincare patch of vacuum AdS. An indication that this is not the right procedure comes from 
Ref.~\cite{Bhattacharyya:2009uu}, whose authors analyzed (mainly) the gravitational collapse of a massless scalar field triggered by turning on its non-normalizable mode with a small amplitude. The naive small amplitude expansion on top of  vacuum AdS had to be resummed so that the background became the AdS-Vaidya solution instead. The latter geometry has a macroscopic horizon which governs the dissipation of perturbations propagating on top of it. This suggests that, in cases of expanding plasmas, the correct procedure is to linearize around some other expanding configuration. We will report elsewhere on studies in this direction in the context of the simplest expanding plasma system, the boost-invariant flow \cite{Bjorken:1982qr}. 

\acknowledgments
We would like to thank D.~Trancanelli for collaboration on the initial stages of this work and G.~Arutyunov, H.~Bantilan, D.~Berenstein, J.~Casalderrey,  P.~Chesler, G.~Horowitz, T.~Jacobson, R.~Janik, J.~Noronha, T.~Peitzmann, F.~Pretorius,  J.~Santos, K.~Schalm, E.~Shuryak, R.~Snellings, C.~Sopuerta, U.~Sperhake and L.~Yaffe for discussions. MPH acknowledges support from the Netherlands Organization for Scientific Research under the NWO Veni scheme and would like to thank Nordita for hospitality while this work was in progress. DM is supported by grants ERC StG HoloLHC-306605, 2009-SGR-168, MEC FPA 2010-20807-C02-01, MEC FPA 2010-20807-C02-02,  CPAN CSD 2007-00042 Consolider-Ingenio 2010. WS acknowledges support from a Utrecht University Foundations of Science grant. For symbolic GR calculations we used M. Headrick's excellent Mathematica package \href{http://people.brandeis.edu/~headrick/Mathematica/index.html}{\tt diffgeo.m}.
\bibliography{iso_biblio}{}

\providecommand{\href}[2]{#2}\begingroup\raggedright\begin{thebibliography}{10}

\bibitem{Heller:2012km}
M.~P. Heller, D.~Mateos, W.~van~der Schee, and D.~Trancanelli, ``{Strong
  Coupling Isotropization of Non-Abelian Plasmas Simplified},''
  \href{http://dx.doi.org/10.1103/PhysRevLett.108.191601}{{\em Phys.Rev.Lett.}
  {\bf 108} (2012)  191601},
\href{http://arxiv.org/abs/1202.0981}{{\tt arXiv:1202.0981 [hep-th]}}.
%%CITATION = ARXIV:1202.0981;%%.

\bibitem{Maldacena:1997re}
J.~M. Maldacena, ``{The Large N limit of superconformal field theories and
  supergravity},'' \href{http://dx.doi.org/10.1023/A:1026654312961}{{\em
  Adv.Theor.Math.Phys.} {\bf 2} (1998)  231--252},
\href{http://arxiv.org/abs/hep-th/9711200}{{\tt arXiv:hep-th/9711200
  [hep-th]}}.
%%CITATION = HEP-TH/9711200;%%.

\bibitem{Witten:1998qj}
E.~Witten, ``{Anti-de Sitter space and holography},'' {\em
  Adv.Theor.Math.Phys.} {\bf 2} (1998)  253--291,
\href{http://arxiv.org/abs/hep-th/9802150}{{\tt arXiv:hep-th/9802150
  [hep-th]}}.
%%CITATION = HEP-TH/9802150;%%.

\bibitem{Gubser:1998bc}
S.~Gubser, I.~R. Klebanov, and A.~M. Polyakov, ``{Gauge theory correlators from
  noncritical string theory},''
  \href{http://dx.doi.org/10.1016/S0370-2693(98)00377-3}{{\em Phys.Lett.} {\bf
  B428} (1998)  105--114},
\href{http://arxiv.org/abs/hep-th/9802109}{{\tt arXiv:hep-th/9802109
  [hep-th]}}.
%%CITATION = HEP-TH/9802109;%%.

\bibitem{Chesler:2008hg}
P.~M. Chesler and L.~G. Yaffe, ``{Horizon formation and far-from-equilibrium
  isotropization in supersymmetric Yang-Mills plasma},''
  \href{http://dx.doi.org/10.1103/PhysRevLett.102.211601}{{\em Phys.Rev.Lett.}
  {\bf 102} (2009)  211601},
\href{http://arxiv.org/abs/0812.2053}{{\tt arXiv:0812.2053 [hep-th]}}.
%%CITATION = ARXIV:0812.2053;%%.

\bibitem{Chesler:2009cy}
P.~M. Chesler and L.~G. Yaffe, ``{Boost invariant flow, black hole formation,
  and far-from-equilibrium dynamics in N = 4 supersymmetric Yang-Mills
  theory},'' \href{http://dx.doi.org/10.1103/PhysRevD.82.026006}{{\em
  Phys.Rev.} {\bf D82} (2010)  026006},
\href{http://arxiv.org/abs/0906.4426}{{\tt arXiv:0906.4426 [hep-th]}}.
%%CITATION = ARXIV:0906.4426;%%.

\bibitem{Chesler:2010bi}
P.~M. Chesler and L.~G. Yaffe, ``{Holography and colliding gravitational shock
  waves in asymptotically AdS$_5$ spacetime},''
  \href{http://dx.doi.org/10.1103/PhysRevLett.106.021601}{{\em Phys.Rev.Lett.}
  {\bf 106} (2011)  021601},
\href{http://arxiv.org/abs/1011.3562}{{\tt arXiv:1011.3562 [hep-th]}}.
%%CITATION = ARXIV:1011.3562;%%.

\bibitem{Heller:2011ju}
M.~P. Heller, R.~A. Janik, and P.~Witaszczyk, ``{The characteristics of
  thermalization of boost-invariant plasma from holography},''
  \href{http://dx.doi.org/10.1103/PhysRevLett.108.201602}{{\em Phys.Rev.Lett.}
  {\bf 108} (2012)  201602},
\href{http://arxiv.org/abs/1103.3452}{{\tt arXiv:1103.3452 [hep-th]}}.
%%CITATION = ARXIV:1103.3452;%%.

\bibitem{Heller:2012je}
M.~P. Heller, R.~A. Janik, and P.~Witaszczyk, ``{A numerical relativity
  approach to the initial value problem in asymptotically Anti-de Sitter
  spacetime for plasma thermalization - an ADM formulation},''
  \href{http://dx.doi.org/10.1103/PhysRevD.85.126002}{{\em Phys.Rev.} {\bf D85}
  (2012)  126002},
\href{http://arxiv.org/abs/1203.0755}{{\tt arXiv:1203.0755 [hep-th]}}.
%%CITATION = ARXIV:1203.0755;%%.

\bibitem{Bantilan:2012vu}
H.~Bantilan, F.~Pretorius, and S.~S. Gubser, ``{Simulation of Asymptotically
  AdS5 Spacetimes with a Generalized Harmonic Evolution Scheme},''
  \href{http://dx.doi.org/10.1103/PhysRevD.85.084038}{{\em Phys.Rev.} {\bf D85}
  (2012)  084038},
\href{http://arxiv.org/abs/1201.2132}{{\tt arXiv:1201.2132 [hep-th]}}.
%%CITATION = ARXIV:1201.2132;%%.

\bibitem{Schee:2013ab}
W.~van~der Schee, ``{Holographic thermalization with radial flow},''
  \href{http://dx.doi.org/10.1103/PhysRevD.87.061901}{{\em Phys.Rev.D.} {\bf
  87} (2013)  061901 (R)}, \href{http://arxiv.org/abs/1211.2218}{{\tt
  arXiv:1211.2218 [hep-th]}}.

\bibitem{Heinz:2004pj}
U.~W. Heinz, ``{Thermalization at RHIC},''
  \href{http://dx.doi.org/10.1063/1.1843595}{{\em AIP Conf.Proc.} {\bf 739}
  (2005)  163--180},
\href{http://arxiv.org/abs/nucl-th/0407067}{{\tt arXiv:nucl-th/0407067
  [nucl-th]}}.
%%CITATION = NUCL-TH/0407067;%%.

\bibitem{CasalderreySolana:2011us}
J.~Casalderrey-Solana, H.~Liu, D.~Mateos, K.~Rajagopal, and U.~A. Wiedemann,
  ``{Gauge/String Duality, Hot QCD and Heavy Ion Collisions},''
\href{http://arxiv.org/abs/1101.0618}{{\tt arXiv:1101.0618 [hep-th]}}.
%%CITATION = ARXIV:1101.0618;%%.

\bibitem{Witten:1998zw}
E.~Witten, ``{Anti-de Sitter space, thermal phase transition, and confinement
  in gauge theories},'' {\em Adv.Theor.Math.Phys.} {\bf 2} (1998)  505--532,
\href{http://arxiv.org/abs/hep-th/9803131}{{\tt arXiv:hep-th/9803131
  [hep-th]}}.
%%CITATION = HEP-TH/9803131;%%.

\bibitem{Bhattacharyya:2008jc}
S.~Bhattacharyya, V.~E. Hubeny, S.~Minwalla, and M.~Rangamani, ``{Nonlinear
  Fluid Dynamics from Gravity},''
  \href{http://dx.doi.org/10.1088/1126-6708/2008/02/045}{{\em JHEP} {\bf 0802}
  (2008)  045},
\href{http://arxiv.org/abs/0712.2456}{{\tt arXiv:0712.2456 [hep-th]}}.
%%CITATION = ARXIV:0712.2456;%%.

\bibitem{Hubeny:2011hd}
V.~E. Hubeny, S.~Minwalla, and M.~Rangamani, ``{The fluid/gravity
  correspondence},''
\href{http://arxiv.org/abs/1107.5780}{{\tt arXiv:1107.5780 [hep-th]}}.
%%CITATION = ARXIV:1107.5780;%%.

\bibitem{AbajoArrastia:2010yt}
J.~Abajo-Arrastia, J.~Aparicio, and E.~Lopez, ``{Holographic Evolution of
  Entanglement Entropy},''
  \href{http://dx.doi.org/10.1007/JHEP11(2010)149}{{\em JHEP} {\bf 1011} (2010)
   149},
\href{http://arxiv.org/abs/1006.4090}{{\tt arXiv:1006.4090 [hep-th]}}.
%%CITATION = ARXIV:1006.4090;%%.

\bibitem{Balasubramanian:2010ce}
V.~Balasubramanian, A.~Bernamonti, J.~de~Boer, N.~Copland, B.~Craps, {\em et
  al.}, ``{Thermalization of Strongly Coupled Field Theories},''
  \href{http://dx.doi.org/10.1103/PhysRevLett.106.191601}{{\em Phys.Rev.Lett.}
  {\bf 106} (2011)  191601},
\href{http://arxiv.org/abs/1012.4753}{{\tt arXiv:1012.4753 [hep-th]}}.
%%CITATION = ARXIV:1012.4753;%%.

\bibitem{Balasubramanian:2011ur}
V.~Balasubramanian, A.~Bernamonti, J.~de~Boer, N.~Copland, B.~Craps, {\em et
  al.}, ``{Holographic Thermalization},''
  \href{http://dx.doi.org/10.1103/PhysRevD.84.026010}{{\em Phys.Rev.} {\bf D84}
  (2011)  026010},
\href{http://arxiv.org/abs/1103.2683}{{\tt arXiv:1103.2683 [hep-th]}}.
%%CITATION = ARXIV:1103.2683;%%.

\bibitem{Aparicio:2011zy}
J.~Aparicio and E.~Lopez, ``{Evolution of Two-Point Functions from
  Holography},'' \href{http://dx.doi.org/10.1007/JHEP12(2011)082}{{\em JHEP}
  {\bf 1112} (2011)  082},
\href{http://arxiv.org/abs/1109.3571}{{\tt arXiv:1109.3571 [hep-th]}}.
%%CITATION = ARXIV:1109.3571;%%.

\bibitem{Bhattacharyya:2009uu}
S.~Bhattacharyya and S.~Minwalla, ``{Weak Field Black Hole Formation in
  Asymptotically AdS Spacetimes},''
  \href{http://dx.doi.org/10.1088/1126-6708/2009/09/034}{{\em JHEP} {\bf 0909}
  (2009)  034},
\href{http://arxiv.org/abs/0904.0464}{{\tt arXiv:0904.0464 [hep-th]}}.
%%CITATION = ARXIV:0904.0464;%%.

\bibitem{Horowitz:1999jd}
G.~T. Horowitz and V.~E. Hubeny, ``{Quasinormal modes of AdS black holes and
  the approach to thermal equilibrium},''
  \href{http://dx.doi.org/10.1103/PhysRevD.62.024027}{{\em Phys.Rev.} {\bf D62}
  (2000)  024027},
\href{http://arxiv.org/abs/hep-th/9909056}{{\tt arXiv:hep-th/9909056
  [hep-th]}}.
%%CITATION = HEP-TH/9909056;%%.

\bibitem{Price:1994pm}
R.~H. Price and J.~Pullin, ``{Colliding black holes: The Close limit},''
  \href{http://dx.doi.org/10.1103/PhysRevLett.72.3297}{{\em Phys.Rev.Lett.}
  {\bf 72} (1994)  3297--3300},
\href{http://arxiv.org/abs/gr-qc/9402039}{{\tt arXiv:gr-qc/9402039 [gr-qc]}}.
%%CITATION = GR-QC/9402039;%%.

\bibitem{Anninos:1995vf}
P.~Anninos, R.~H. Price, J.~Pullin, E.~Seidel, and W.-M. Suen, ``{Headon
  collision of two black holes: Comparison of different approaches},''
  \href{http://dx.doi.org/10.1103/PhysRevD.52.4462}{{\em Phys.Rev.} {\bf D52}
  (1995)  4462--4480},
\href{http://arxiv.org/abs/gr-qc/9505042}{{\tt arXiv:gr-qc/9505042 [gr-qc]}}.
%%CITATION = GR-QC/9505042;%%.

\bibitem{'tHooft:1973jz}
G.~'t~Hooft, ``{A Planar Diagram Theory for Strong Interactions},''
\href{http://dx.doi.org/10.1016/0550-3213(74)90154-0}{{\em Nucl.Phys.} {\bf
  B72} (1974)  461}.
%%CITATION = NUPHA,B72,461;%%.

\bibitem{Mukhopadhyay:2012hv}
A.~Mukhopadhyay, ``{Non-equilibrium fluctuation-dissipation relation from
  holography},'' \href{http://dx.doi.org/10.1103/PhysRevD.87.066004}{{\em
  Phys.Rev.} {\bf D87} (2013)  066004},
\href{http://arxiv.org/abs/1206.3311}{{\tt arXiv:1206.3311 [hep-th]}}.
%%CITATION = ARXIV:1206.3311;%%.

\bibitem{Bhattacharyya:2008mz}
S.~Bhattacharyya, R.~Loganayagam, I.~Mandal, S.~Minwalla, and A.~Sharma,
  ``{Conformal Nonlinear Fluid Dynamics from Gravity in Arbitrary
  Dimensions},'' \href{http://dx.doi.org/10.1088/1126-6708/2008/12/116}{{\em
  JHEP} {\bf 0812} (2008)  116},
\href{http://arxiv.org/abs/0809.4272}{{\tt arXiv:0809.4272 [hep-th]}}.
%%CITATION = ARXIV:0809.4272;%%.

\bibitem{Janik:2008tc}
R.~A. Janik and P.~Witaszczyk, ``{Towards the description of anisotropic plasma
  at strong coupling},''
  \href{http://dx.doi.org/10.1088/1126-6708/2008/09/026}{{\em JHEP} {\bf 0809}
  (2008)  026},
\href{http://arxiv.org/abs/0806.2141}{{\tt arXiv:0806.2141 [hep-th]}}.
%%CITATION = ARXIV:0806.2141;%%.

\bibitem{deHaro:2000xn}
S.~de~Haro, S.~N. Solodukhin, and K.~Skenderis, ``{Holographic reconstruction
  of space-time and renormalization in the AdS / CFT correspondence},''
  \href{http://dx.doi.org/10.1007/s002200100381}{{\em Commun.Math.Phys.} {\bf
  217} (2001)  595--622},
\href{http://arxiv.org/abs/hep-th/0002230}{{\tt arXiv:hep-th/0002230
  [hep-th]}}.
%%CITATION = HEP-TH/0002230;%%.

\bibitem{Beuf:2009cx}
G.~Beuf, M.~P. Heller, R.~A. Janik, and R.~Peschanski, ``{Boost-invariant early
  time dynamics from AdS/CFT},''
  \href{http://dx.doi.org/10.1088/1126-6708/2009/10/043}{{\em JHEP} {\bf 0910}
  (2009)  043},
\href{http://arxiv.org/abs/0906.4423}{{\tt arXiv:0906.4423 [hep-th]}}.
%%CITATION = ARXIV:0906.4423;%%.

\bibitem{Wu:2011ab}
B.~Wu and P.~Romatschke, ``{Shock wave collisions in AdS5: approximate
  numerical solutions},''
  \href{http://dx.doi.org/10.1142/S0129183111016920}{{\em Int. J. of Mod. Phys.
  C} {\bf 22} (2011)  1317--1342}, \href{http://arxiv.org/abs/1108.3715}{{\tt
  arXiv:1108.3715 [hep-th]}}.

\bibitem{lrr-2009-1}
P.~Grandclément and J.~Novak, ``Spectral methods for numerical relativity,''
  {\em Living Reviews in Relativity} {\bf 12} (2009) no.~1, .
  \url{http://www.livingreviews.org/lrr-2009-1}.

\bibitem{Hartnoll:2009sz}
S.~A. Hartnoll, ``{Lectures on holographic methods for condensed matter
  physics},'' \href{http://dx.doi.org/10.1088/0264-9381/26/22/224002}{{\em
  Class.Quant.Grav.} {\bf 26} (2009)  224002},
\href{http://arxiv.org/abs/0903.3246}{{\tt arXiv:0903.3246 [hep-th]}}.
%%CITATION = ARXIV:0903.3246;%%.

\bibitem{lrr-1999-2}
K.~D. Kokkotas and B.~Schmidt, ``Quasi-normal modes of stars and black holes,''
  {\em Living Reviews in Relativity} {\bf 2} (1999) no.~2, .
  \url{http://www.livingreviews.org/lrr-1999-2}.

\bibitem{Mateos:2011bs}
D.~Mateos, ``{Gauge/string duality applied to heavy ion collisions:
  Limitations, insights and prospects},''
  \href{http://dx.doi.org/10.1088/0954-3899/38/12/124030}{{\em J.Phys.} {\bf
  G38} (2011)  124030},
\href{http://arxiv.org/abs/1106.3295}{{\tt arXiv:1106.3295 [hep-th]}}.
%%CITATION = ARXIV:1106.3295;%%.

\bibitem{Bjorken:1982qr}
J.~Bjorken, ``{Highly Relativistic Nucleus-Nucleus Collisions: The Central
  Rapidity Region},''
\href{http://dx.doi.org/10.1103/PhysRevD.27.140}{{\em Phys.Rev.} {\bf D27}
  (1983)  140--151}.
%%CITATION = PHRVA,D27,140;%%.

\end{thebibliography}\endgroup
\bibliographystyle{utphys}
\end{document}